\documentclass[preprint2]{aa}
\usepackage{graphicx}
\usepackage{txfonts}
\newcommand{\1}{{~\sc i}}
\newcommand{\2}{{~\sc ii}}
\newcommand{\3}{{~\sc iii}}
\newcommand{\4}{{~\sc iv}}
\newcommand{\5}{{~\sc v}}
\newcommand{\6}{{~\sc vi}}
\newcommand{\kms}{{\,km\,s$^{-1}$}}

\begin{document}

   \title{Metal enrichment of the neutral gas of blue compact dwarf galaxies: the compelling case of Pox\,36}

   \author{V. Lebouteiller\inst{1},
          D. Kunth\inst{2},
          T.X. Thuan\inst{3},
          \and
          J.M. D\'esert\inst{2}
          }

   \institute{1: Center for Radiophysics and Space Research, Cornell University, Space Sciences Building, Ithaca, NY 14853-6801   \\
   2: Institut d'Astrophysique, Paris, 98 bis Boulevard Arago, F-75014 Paris \\
  3: Astronomy Department, University of Virginia, Charlottesville, VA 22903 \\
            \email{vianney@isc.astro.cornell.edu}
             }

   \date{}

\abstract
{Evidence has grown over the past few years that the neutral phase of blue compact dwarf (BCD) galaxies may be metal-deficient as compared to the ionized gas of their H\2\ regions. These results have strong implications for our understanding of the chemical evolution of galaxies, and it is essential to strengthen the method, as well as to find possible explanations. }
{We present the analysis of the interstellar spectrum of Pox\,36 with the \textit{Far Ultraviolet Spectroscopic Explorer} (\textit{FUSE}).  Pox\,36 was selected because of the relatively low foreground gas content that makes it possible to detect absorption-lines weak enough that unseen components should not be saturated.}
{Interstellar lines of H\1, N\1, O\1, Si\2, P\2, Ar\1, and Fe\2\ are detected. Column densities are derived directly from the observed line profiles except for H\1, whose lines are contaminated by stellar absorption, thus needing the stellar continuum to be removed. We used the TLUSTY models to remove the stellar continuum and isolate the interstellar component. The best fit indicates that the dominant stellar population is B0. The observed far-UV flux agrees with an equivalent number of $\sim300$ B0 stars. The fit of the interstellar H\1\ line gives a column density of $10^{20.3\pm0.4}$\,cm$^{-2}$. Chemical abundances were then computed from the column densities using the dominant ionization stage in the neutral gas. Our abundances are compared to those measured from emission-line spectra in the optical, probing the ionized gas of the H\2\ regions.}
{Our results suggest that the neutral gas of Pox\,36 is metal-deficient by a factor $\sim7$ as compared to the ionized gas, and they agree with a metallicity of $\approx1/35$\,Z$_\odot$. Elemental depletion is not problematic because of the low dust content along the selected lines of sight. In contrast, the ionized gas shows a clear depletion pattern, with iron being strongly depleted. }
{The abundance discontinuity between the neutral and ionized phases implies that most of the metals released by consecutive star-formation episodes mixes with the H\1\ gas. The volume extent of the enrichment is so large that the metallicity of the neutral gas increases only slightly. The star-forming regions could be enriched only by a small fraction ($\sim1\%$), but it would greatly enhance its metallicity. Our results are compared to those of other BCDs. We confirm the overall underabundance of metals in their neutral gas, with perhaps only the lowest metallicity BCDs showing no discontinuity. }

\keywords{ISM: abundances, (ISM): HII regions, line: profiles, galaxies: abundances, galaxies: dwarf, galaxies: individual: Pox 36}

\authorrunning{V. Lebouteiller et al.}

\maketitle

\section{Introduction}

In the chemical downsizing scenario, in which massive galaxies are the first to form stars at a high rate (e.g., Cen \&\ Ostriker 1999), dwarf galaxies can remain chemically unevolved until later times. It is thus theoretically possible that some dwarf galaxies in the nearby Universe still contain pristine gas. The first observational constraint was provided by studies of chemical abundances in the ionized gas of blue compact dwarf galaxies (BCDs). These studies yield a fundamental result: a metallicity floor appears to exist, with no BCD having an ionized gas oxygen abundance lower than $12+\log {\rm (O/H)} \sim 7.0$, or Z$_\odot$/50\footnote{We use hereafter the solar oxygen abundance $12+\log {\rm (O/H)} = 8.66$ of Asplund et al.\ (2005)}. For more than three decades, I Zw 18 discovered by Zwicky (1966) and first studied by Sargent \& Searle (1970) held the record as the most metal-deficient galaxy known, with an oxygen abundance $12+\log {\rm (O/H)} = 7.17\pm0.01$ in its NW component and $7.22\pm0.02$ in its SE component (Thuan \& Izotov 2005). Only very recently has I\,Zw\,18 been displaced by two BCDs that are more metal-deficient, SBS\,0335--052W with $12+\log {\rm (O/H)} = 7.12\pm0.03$ (Izotov et al.\ 2005) and DDO\,68 with $12+\log {\rm (O/H)} = 7.14\pm0.03$ (Izotov \& Thuan 2007).
The difficulty of finding more metal-poor objects was originally explained by the local mixing of the amount of metals produced and released in 
a single starburst episode (Kunth \& Sargent 1986). This hypothesis has been since challenged by observations of uniform abundances in disconnected star-forming regions within a single dwarf galaxy (e.g., Skillman \& Kennicut 1993; Kobulnicky \& Skillman 1997; Noeske et al.\ 2000). 

Tenorio-Tagle (1996) proposed an attractive scenario in which supernovae (SNe) products are driven to high galactic latitudes in a hot phase ($\sim10^6$\,K) before they cool down, condense as molecular droplets, and rain back on the disk. The complete cycle (before mixing) would last several $\sim100$\,Myr. This scenario naturally explains the uniform abundances found in dwarf galaxies, while ruling out local enrichment within the star-forming regions. It must be noted that models of Rieschick \& Hensler (2001; 2003), based on Tenorio-Tagle's scenario, predict that only $\sim75$\%\ of the SNe products go through the outer galactic cycle as hot gas, while the remaining 25\%\ evolve in an inner local cycle and can enrich the star-forming regions in only 10\,Myr. A few percent could eventually leave the gravitational field of the galaxy. On the other hand, {\it Chandra} X-ray observations of SBS\,0335--052W, SBS\,0335--052E, and I\,Zw\,18 do not reveal any sign of a hot gas breaking out from the stellar body implying that the galactic fountain scenario may not be valid for BCDs (Thuan et al.\ 2004). In fact, Recchi et al.\ (2001; 2004) modeled star-formation in gas-rich dwarf galaxies and found that the majority of newly produced metals resides in a cold phase ($\lesssim2\times10^4$\,K) after only a few tens of Myr, supporting the common instantaneous mixing hypothesis. 

To understand the metal enrichment of galaxies, it is necessary to study all the gaseous phases involved in the gas mixing cycle.
X-ray observations of starburst galaxies led to the conclusion that the hot gas is probably enriched by SNe products (e.g., Martin et al.\ 2002). The following step in the mixing cycle is poorly constrained observationally. Do metals mix with the cold halo and thus enriching the H\1\ region? Do they eventually condense as molecular droplets and settle back onto the stellar body? The situation might be different in BCDs if galactic fountains do not form. However, it seems that winds from massive stars and SNe are still able to carve the ISM to form superbubbles in which metals are injected at high temperatures (Thuan et al.\ 2004). Mixing could then occur at the interfaces between the hot and cold gas. Thus the study of the metal content in the neutral gas of dwarf galaxies provides us with essential constraints for their chemical evolution models. 

The neutral phase usually consists of at least two distinct components (Field et al.\ 1969), with a warm medium (WNM; $T_w\sim10^4$\,K, $n_w\sim$ few tenths cm$^{-3}$) showing broad hyperfine 21\,cm line emission, and a cold medium (CNM, $T_c\sim100$\,K, $n_c\sim$ few tens cm$^{-3}$) showing narrow 21\,cm emission and absorption. It must be noted that the distinction between CNM and WNM becomes indistinct in galaxies with low-pressure ISM, such as dwarfs (Lo et al.\ 1993). Neutral metals can be observed through resonant lines in the far-ultraviolet (FUV). BCDs are ideal targets because they display large amounts of H\1\ gas (Thuan \& Martin 1981), and because the massive stars provide a strong FUV continuum. Kunth et al.\ (1994) used the \textit{GHRS} instrument on board the \textit{Hubble Space Telescope} to observe interstellar absorption lines along the lines of sight toward the many massive stars in IZw18. The authors detected the O\1\ line at 1302.2\,\AA\ and, together with the neutral hydrogen content measured from the 21\,cm line, estimated the metallicity of the neutral envelope to be at least 10 times lower than that of the ionized gas in the H\2\ regions. Later, Thuan \& Izotov (1997) obtained a \textit{GHRS} spectrum for SBS\,0335--052E and found the metallicity of the neutral gas to be similar to that of the ionized gas, about 2\% solar.  These results remained however inconclusive because of possible saturation effects of the O\1\ absorption line (see Pettini \& Lipman 1995). 

A new step forward was achieved with the launch of the \textit{Far Ultraviolet Spectroscopic Explorer} (\textit{FUSE}; Moos et al.\ 2000), which allows the observation of absorption-lines of H\1\ together with many metallic species such as N\1, O\1, Si\2, P\2, Ar\1, and Fe\2. The spectral resolution of \textit{FUSE} ($R\sim20\,000$) and its sensitivity made it possible to detect numerous absorption-lines, some of them apparently not saturated. The neutral gas chemical composition was derived in several BCDs, spanning a wide range in ionized gas metallicity, from 1/50 to 1/3 solar, with I\,Zw\,18 (Aloisi et al.\ 2003; Lecavelier des Etangs et al.\ 2004), SBS\,0335--052 (Thuan et al.\ 2005), I\,Zw\,36 (Lebouteiller et al.\ 2004), Mark\,59 (Thuan et al.\ 2002), NGC\,625 (Cannon et al.\ 2005), and NGC\,1705 (Heckman et al.\ 2001). These investigations showed that the neutral gas of BCDs is not pristine, but that it has already been enriched with metals up to an amount of $\gtrsim1/50$\,Z$_\odot$. The second most important result is that the metallicity of the neutral gas is generally lower than that of the ionized gas $-$ except for the two lowest-metallicity BCDs, I\,Zw\,18 and SBS\,0335--052E $-$ implying that the neutral phase has been probably less processed. 

The interpretation of these remarkable results has been limited so far by the lack of knowledge about the origin of the neutral gas in BCDs whose absorption is detected in the FUV. Direct comparison with the H\1\ gas seen \textit{via} the 21\,cm emission is hampered by a bias in selecting out relatively dust-free FUV lines of sight toward the massive stars. Even more importantly, FUV observations only probe gas in front of stellar clusters while radio observations in emission probe the whole system. Many uncertainties also remain concerning the method of deriving column densities from absorption-lines in complex systems. Most \textit{FUSE} studies of BCDs assumed a single homogeneous line of sight with the exception of Mark\,59 (Thuan et al.\ 2002; see also Lebouteiller et al.\ 2006 for the giant H\2\ region NGC\,604). A more realistic approach would be to consider multiple lines of sight toward massive stars, each line of sight intersecting clouds with possibly different physical conditions (such as turbulent velocity $b$, radial velocity $v$) and chemical properties. If individual absorption components are unresolved, it is almost impossible to tell whether they are saturated, in which case the column density inferred from the global absorption line can be severely underestimated (see e.g., Lebouteiller et al.\ 2006). Following the suggestion of Kunth \&\ Sargent (1986), 
Bowen et al.\ (2005) got around the problem by deriving the abundances in the neutral gas of the dwarf spiral galaxy SBS\,1543+593 along a quasar line of sight. The authors found that the abundances agree with those in the ionized gas of the brightest H\2\ region in the arms, contrasting with the \textit{FUSE} results in BCDs. As far as chemical models are concerned, Recchi et al.\ (2004) examined the case of I\,Zw\,18 for which they found that abundances in the H\1\ medium (defined by a temperature lower than $7\,000$\,K) are similar to those in the H\2\ gas. Hence, although some physics is still missing in the models (radiative transfer, clumpy ISM, dust phase), it seems that \textit{FUSE} results still cannot be reproduced. 

In this paper, we look anew at the problem of how the chemical compositions of the ionized and the neutral gas in BCDs compare by selecting an object with an extremely low H\1\ column density. This object has some absorption lines that are weak enough to be safely considered as unsaturated. The object we analyze is Pox\,36 (also known as ESO\,572--G\,034, I\,SZ\,63, HIPASSJ1158-19b, or IRAS\,11564--1844), an emission-line galaxy first studied by Rodgers et al.\ (1978) and then rediscovered by Kunth et al.\ (1981). It is characterized by strong optical emission-lines superimposed on a very blue continuum.
Its spectral characteristics and morphological properties fit the definition of a BCD galaxy (see e.g., Thuan \& Martin 1981, Kunth \& \"Ostlin 2000).
Its known observational properties are listed in Table\,\ref{tab:prop}. In particular, chemical abundances in the ionized gas of the H\2\ regions within Pox\,36 have been derived by several independent studies (e.g., Kunth \& Sargent 1983; Izotov \& Thuan 2004). The ionized gas has an oxygen abundance $12+\log {\rm (O/H)} \approx 8.05$ which corresponds to Z$_\odot$/5. With a H\1\ velocity of 1114 km\,s$^{-1}$ (Meyer et al.\ 2004), a Hubble constant of 73 km\,s$^{-1}$\,Mpc$^{-1}$, its cosmology-corrected distance is 20\,Mpc.  

The {\it FUSE} observations are presented in Sect.\,\ref{sec:observations} and an overview of the \textit{FUSE} spectrum is given in Sect.\,\ref{sec:overview}.
We infer the column densities of H\1\ and of the metals from the line profile in Sect.\,\ref{sec:fitting}. Chemical abundances are derived and discussed in Sect.\,\ref{sec:abundances}. Finally, we study the properties of the neutral gas in the current {\it FUSE} BCD sample to discuss general trends in Sect.\,\ref{sec:global}.

\section{Observations}\label{sec:observations}

\begin{figure}[b!]
\centering
\includegraphics[angle=0,scale=0.45,clip=true]{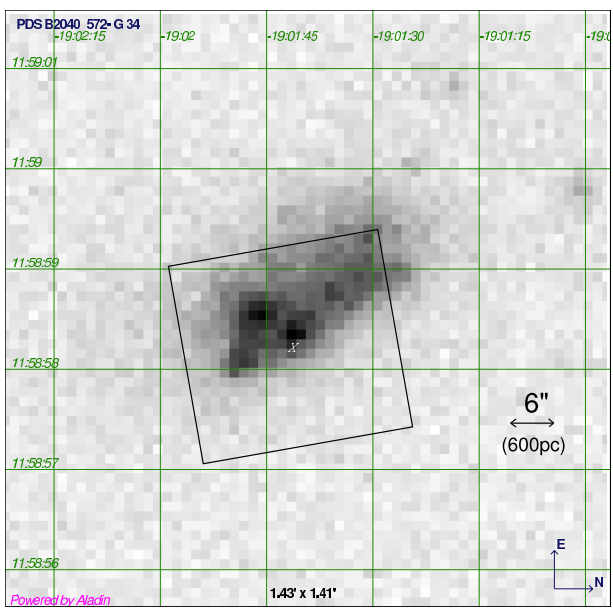}
\caption{Blue band image of Pox\,36 (Astro-Wise, ESO-LV Catalog). The separation between the 2 bright knots is given by the segment. The inclined square indicates the $30''\times30''$ LWRS aperture. 
 \label{fig:pox36}}
\end{figure}

Pox\,36 was observed by the \textit{FUSE} telescope on May 11 2001 (observation B0220201). All eight observing channels were used with the LWRS aperture ($30''\times30''$). At the distance of Pox\,36 (see Table\,\ref{tab:prop}), $1''$ corresponds to $\approx100$\,pc. The total exposure time is 25\,470\,s. Data were taken in "Time Tag" observing mode. The brightest blue regions are included in the aperture as well as the fainter extended emission seen toward the north-east (Fig.\,\ref{fig:pox36}). In the blue band, the combined emission from the bright knots is on the same order as the emission from the rest of the body. We expect shorter wavelength light such as the FUV emission to be dominated by the knots where most of the massive stars lie.

\begin{table*}
\caption{Observational properties of Pox\,36.}
\label{tab:prop}
\begin{tabular}{l l l }
\hline\hline
Property     &  Value         &   Comment          \\ 
\hline  
$\alpha$ (J2000) & $11{\rm h}58{\rm m}57.3{\rm s}$ &           HIPASS (Meyer et al.\ 2004) \\
$\delta$ (J2000) & $-19^\circ01'41''$ &           HIPASS (Meyer et al.\ 2004) \\
Galactic coordinates & $l=286.114$, $b=42.123$ &              HIPASS (Meyer et al.\ 2004) \\
\hline
Foreground Galactic extinction (mag) & $E(B-V)=0.038$  & Schlegel et al.\ (1998)  \\ 
\hline
Magnitude (mag) & $B_J=14.27$, $I=14.38$  & HIPASS (Doyle et al.\ 2005) \\
                    & $K_s=12.63$                     & Vanzi \& Sauvage (2006) \\
\hline
L(H$\beta$) (ergs s$^{-1}$) &  $1.81\times10^{39}$   & From F(H$\beta$) (Izotov \& Thuan 2004) and D = 20\,Mpc \\
\hline
FWHM (H\1)  (\kms) & $67.4$ &                     HIPASS (Meyer et al.\ 2004) \\
Peak H\1\ flux (Jy) & $0.1197$ &                     HIPASS (Meyer et al.\ 2004) \\
Integrated H\1\ flux (Jy\kms) & $8.6$ &                      HIPASS (Meyer et al.\ 2004) \\
$M$(H\1) (M$_\odot$)     & $8.1\times10^8$                &   Derived from the H\1\ flux with the distance=$20$\,Mpc \\
\hline
Radial velocity (\kms)   &    $1114\pm2$  &   21\,cm line, Theureau et al.\ (1998) \\ 
                               & $1113.9\pm6.5$ &                      HIPASS (Meyer et al.\ 2004; Zwaan et al.\ 2004) \\
& $1065\pm20$ & Optical lines (Izotov \& Thuan 2004) \\
\hline
Distance (Mpc)   & $\approx12$           &    Vigroux et al.\ (1987) with H$_0=73$\kms  \\
	         & $\approx16$      &    Kunth \& Joubert (1985)\\
	         & $\approx20$      &    Using $v=1114$\kms, H$_0=73$\kms, $\Omega_m=0.27$   \\
\hline
$\log$ O/H  & $-3.95\pm0.02$&                  Izotov \& Thuan (2004) \\ 
     & $-3.86\pm0.03$ &                    Kobulnicky \& Skillman (1996) \\
     & $-3.81\pm0.04$ &                    Campbell (1988; 1992) \\
     & $-3.82\pm0.04$ &                    Kunth \& Sargent (1983) \\
$\log$ N/H  & $-5.53\pm0.03$ &                    Izotov \& Thuan (2004) \\ 
     & $-5.39\pm0.10$ &                     Kobulnicky \&  Skillman (1996) \\
     & $-5.38\pm0.05$ &                      Campbell (1992)   \\
     & $-5.37\pm0.05$ &                    Kunth \& Sargent (1983) \\
$\log$ Ar/H & $-6.24\pm0.03$ &                     Izotov \& Thuan (2004) \\ 
$\log$ Fe/H & $-6.17\pm0.10$ &                     Izotov \& Thuan (2004) \\ 
\hline
\end{tabular}
\end{table*}

\begin{figure*}
\centering
\includegraphics[angle=0,scale=0.65,clip=true]{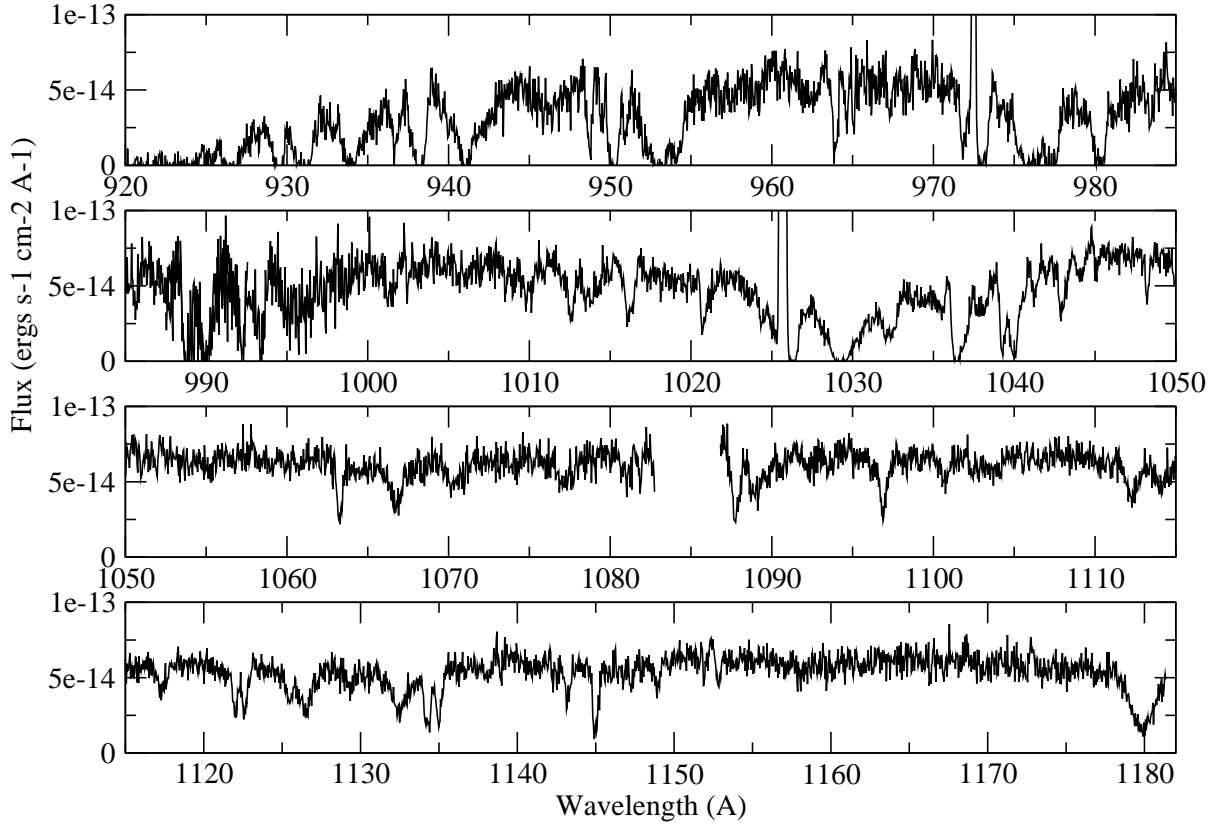}
\caption{Complete \textit{FUSE} spectrum of Pox\,36. The flux is plotted against the observed wavelength (3.5-4\,\AA\ shift from the rest wavelength, Sect.\,\ref{sec:ismcomp}). Data from overlapping channels were co-added for display purposes. The strong emission-lines at $\sim972$\,\AA\ and $\sim1026$\,\AA\ correspond to terrestrial H\1\ airglows.
\label{fig:spectrum}}
\end{figure*}

Data were processed by the CalFUSE\footnote{The CalFUSE calibration pipeline is available at \textit{http://fuse.pha.jhu.edu/analysis/calfuse.html}} pipeline (version 3.2.1). We aligned the individual exposures on the wavelength grid and co-added them to obtain one spectrum per observing channel.
The final spectrum is presented in Fig.\,\ref{fig:spectrum}, in which observations from different channels have been co-added for display purposes. The signal-to-noise ratio per resolution element (20\kms) is $\approx$12. 

The intrinsic spectroscopic resolving power of \textit{FUSE} is close to $R=20\,000$ corresponding to a line spread function (LSF) width of $\sim15$\kms\ (e.g., Sahnow et al.\ 2000; H{\'e}brard et al.\ 2002; Kruk et al.\ 2002).  The effective resolution depends on the quality of the individual exposure alignment. In particular, the alignment is difficult to assess for spectra in the SiC channels because of the smaller effective area as compared to the LiF channels, resulting in a lower signal-to-noise ratio. Misalignments reduce significantly the spectral resolution of the final spectrum so that we estimate the LSF width to be as large as $\sim25$\kms.

The spatial extent of the source within the LWRS aperture can also result in a smearing of the spectrum (see e.g., Lebouteiller et al.\ 2004; 2006). Unfortunately, Pox\,36 has not been imaged in the UV. The optical blue image of Pox\,36 shows two bright knots separated by $\approx6''$ (Fig.\,\ref{fig:pox36}), corresponding to a projected distance of $\approx4''$ in the dispersion direction. Given the LWRS aperture size of $30''=120$\kms, the separation between the 2 knots should result in a wavelength smearing of $\sim16$\kms\ at most. 

Since the line broadening also depends on several observational parameters (such as the pointing jitter), it is difficult to derive \textit{a priori} a precise estimate. We empirically determined the LSF width by using the separation between closed or blended absorption lines. The most likely value is $\sim30$\kms\ for the LiF channels and $\sim37.5$\kms\ for the SiC channels. This is roughly consistent with the quadratic sum of the intrinsic PSF and the smearing due to the spatial extent. This is the LSF width we use for line fitting (Sect.\,\ref{sec:metals}). In practice, our results depend little on the LSF width determination (see Sect.\,\ref{sec:single}).

\section{Overview of the FUSE spectrum}\label{sec:overview}

\subsection{Absorption systems}\label{sec:ismcomp}

The spectral continuum is provided by the UV-bright massive stars in the galaxy. Absorption lines from species along the line of sight are superimposed on the continuum. The line of sight intersects the ISM from the Milky Way and from Pox\,36 itself. As noted by Wakker et al.\ (2003), the line of sight passes at only 18$'$ ($\approx100$\,kpc projected) from NGC\,4027, but there is no absorption visible at the velocity of the latter ($\approx1460$\kms). The redshift of Pox\,36 makes it possible to separate easily the local absorption system from the one in the BCD. We identify absorption lines from the Milky Way at nearly zero radial velocity. We barely detect Galactic H$_2$, which is mostly due to the high Galactic latitude of Pox\,36 (Table 1).

The absorption system corresponding to the neutral ISM in Pox\,36 is detected at a velocity of $v_n \sim 1052$\kms\ ($1058\pm10$\kms\ from the metal lines,  $1047\pm10$\kms\ from the H\1\ lines). This corresponds to a wavelength shift of $\sim3.5-4.0$\,\AA\ from the rest-frame. Only the strongest atomic lines are detected. No H$_2$ line at the redshift of  Pox\,36 is seen. The velocity $v_n$ is significantly smaller than the one derived from the 21\,cm H\1\ line equal to $1113.9\pm6.5$\,km s$^{-1}$ (Meyer et al.\ 2004; Zwaan et al.\ 2004).
However, it must be stressed that the regions probed in the UV and the radio are different in terms of extent (the radio beam size is $15.5'$ as compared to the $30''\times30''$ aperture of the LWRS) and of depth (because of dust extinction).
The aperture (and depth) effect explanation is supported by the fact that $v_n$ is in agreement with the value of 1065$\pm$20 km s$^{-1}$ derived by Izotov \& Thuan (2004) from the optical emission lines of  the ionized gas (Table\,\ref{tab:prop}).
Taking at face the velocity of the neutral and ionized gas tracers, we conclude that there is no significant evidence that the neutral gas is outflowing from the star-forming regions.
 
We find that the turbulent velocity of the neutral gas in Pox\,36 is $b_n = 33\pm10$\kms. Its value is mostly constrained by lines near saturation such as the O\1\ 988.77\,\AA\ line (Table\,\ref{tab:lines}). The quoted uncertainty on the $b$-value does not include possible systematic errors on the LSF width. The corresponding FWHM is 
is $2 b_n \sqrt{\ln 2} \approx 55\pm17$\kms, consistent with the FWHM measured in radio (67.4\kms, Table\,\ref{tab:prop}). 
However as said before, the region probed by the large radio beam ($15.5'$) is not directly comparable to the gas probed in the FUV (corresponding to an extent of less than $\sim10''$, Fig.\,\ref{fig:pox36}). 
It must be also stressed that our determination of $b_n$ may not be physical if the absorption lines arise from a collection of interstellar clouds (see Sects.\,\ref{sec:morpho} and \ref{sec:multi}).

\begin{table*}
\centering
\begin{minipage}[t]{2.00\columnwidth}
\caption{Wavelengths and oscillator strengths of the lines analyzed.}
\label{tab:lines}\renewcommand{\footnoterule}{} 
\begin{tabular}{p{0.2\linewidth} l l l l p{0.2\linewidth}}
\hline\hline
Specie & $\lambda_{\rm rest}$ (\AA) & $\lambda_{\rm obs}$ (\AA) & $\log(gf)$\footnote{Atomic data are from Morton et al.\ (2003), except for Fe\2\ $\lambda$1144.938 for which we take Howk et al.\ (2000).}   & $N_{\rm sat}$ (cm$^{-2}$)\footnote{$N_{\rm sat}$ is the maximum column density before saturation of a central component with $b=2$\kms\ (Sec.\,\ref{sec:multi}).} & Comment \\
\hline  
H\1\      & 1025.722                        & 1029.389                      &  0.0527 &      ...                  & Lyman $\beta$ \\
H\1\     &  937.803                           & 941.080                       & 0.0026  &     ...                   & Lyman   $\epsilon$  \\
\hline
N\1\      &  953.415                         & 956.797                         & 0.0132 &    $5\times10^{14}$ &                      \\ 
N\1\      &  953.655                         & 957.031                         & 0.0250    &   $3\times10^{14}$  &                      \\ 
N\1\      &  953.970                         & 957.343                         & 0.0350    &     $2\times10^{14}$  &                   \\  
N\1\     & 963.990                          & 967.040                          & 0.0148  &  $4\times10^{14}$   &  Blended with P\2\ $\lambda963.8$                     \\ 
N\1\    &  964.626                           &  968.029                       & 0.0094  &  $7\times10^{14}$  &   Barely detected                  \\ 
N\1\     &  965.041                         & 968.458                         & 0.0040  & $2\times10^{15}$   & Not detected \\ 
N\1\     & 1134.165                         & 1138.069                      & 0.0134 & $4\times10^{14}$   &   Barely detected     \\ 
N\1\     &  1134.415                        & 1138.303                     & 0.0268  &  $2\times10^{14}$   &                       \\ 
N\1\     &  1134.980                        & 1138.888                     & 0.0402   & $8\times10^{13}$   &                        \\
\hline
O\1\     & 988.773                           & 992.209                         & 0.0465  & $1\times10^{14}$  & Saturated   \\ 
O\1\    &  1039.230                         & 1042.961                       & 0.0092  & $5\times10^{14}$  &           \\  
\hline
Si\2\   &  1020.699                         &  1024.358                     & 0.0164   & $2\times10^{14}$   &                        \\ 
\hline
P\2\    &  963.800                           & 967.249                         & 1.4600 &$2\times10^{12}$  &    Blended with N\1\  $\lambda963.99$    \\ 
P\2\     & 1152.818                        &   1156.906                     & 0.2450 &   $2\times10^{13}$ &   Not detected  \\ 
\hline
Ar\1\    &  1048.220                       & 1051.957                       & 0.2630   & $9\times10^{12}$  &                          \\ 
\hline
Fe\2\   & 1144.938                        & 1148.976                      & 0.1060  &  $3\times10^{13}$  &      \\ 
\hline
\end{tabular}
\end{minipage}
\end{table*}

\subsection{Neutral ISM components}\label{sec:morpho}

The lines of sight observed by \textit{FUSE} are optically thick to the Lyman continuum of the stellar clusters but the UV radiation is still able to reach us. This implies that the amount of foreground dust is either very small and/or it is clumpy, as expected for metal-poor objects (e.g., Lisenfeld \& Ferrara 1998; Hirashita et al.\ 2002). 
Although \textit{FUSE} observations are sensitive to both the CNM and the WNM as defined in McKee \& Ostriker (1977) and Wolfire et al.\ (1995), the situation might be complex in BCDs since the relatively low ISM pressure could lead to a single H\1\ phase, except in regions affected by stellar winds and supernov\ae\ shocks. Unfortunately, the spectral resolution of \textit{FUSE} (Sect.\,\ref{sec:observations}) is too small to isolate warm and cold neutral phases. 

In principle, the presence of molecular absorption could allow us to constrain the physical conditions along the observed lines of sight. No diffuse H$_2$ has been detected so far along lines of sight toward massive stars in BCDs (e.g., Vidal-Madjar et al.\ 2000). The reasons are intrinsic to the H\1\ phase since H$_2$ has been detected in emission in the warm ISM of these objects (e.g., Vanzi et al.\ 2005; Thuan et al.\ 2005).
Warm H$_2$ has been since observed in several other BCDs by the \textit{Infrared Spectrograph} onboard \textit{Spitzer} (Lebouteiller et al.\ in preparation). The absence of H$_2$ in the diffuse ISM of BCDs is due to the combined effects of a low H\1\ density in the neutral gas envelope, a large UV flux that destroys H$_2$ molecules and a low metallicity that makes grains on which to form H$_2$ molecules scarce (Vidal-Madjar et al.\ 2000). Diffuse H$_2$ is detected in relatively more metal-rich star-forming objects, but the molecular fraction is very low, on the order of $10^{-4}$ (Hoopes et al.\ 2004). A small velocity offset ($\sim10$\,\kms) between H$_2$ lines and atomic lines has been found in NGC\,625 (Cannon et al.\ 2005) and toward the giant H\2\ region NGC\,604 (Lebouteiller et al.\ 2006). Although the offset is small, this might indicate that atomic absorption-lines preferentially arise in regions where H$_2$ is not able to form. Hence, it seems reasonable to assume that the neutral phase probed by FUV absorption-lines is rather warm and diffuse since dense clouds should show H$_2$ absorption.
In the following, we thus assume that the interstellar absorption-lines arise from a single phase dominated by turbulent motions. Since the filling factor is unknown, this phase could be a uniform H\1\ envelope or a collection of H\1\ clouds.

\subsection{Stellar contamination}\label{sec:stellar_cont}

The FUV spectral continuum of most of the BCDs and giant H\2\ regions studied so far with \textit{FUSE} is dominated by O stars. This is indicative of a starburst age younger than $\sim10$\,Myr (Robert et al.\ 2003). O stars have relatively "clean" spectra, the most prominent stellar absorption lines being due to He\2, O\2-\3, and N\2-\3\ in the optical, and to O\6, N\5, P\5, S\4, C\4\ and Si\4\ in the FUV. Because of the high temperature, hydrogen is mostly ionized, and only weak H\1\ photospheric lines can be observed which contribute to the $-$ already saturated $-$ core of the interstellar H\1\ absorption line. Hence FUV spectra toward young starbursts allow a precise determination of the interstellar H\1\ content. The only exception concerns extremely young starburst episodes showing a prominent O\6\ P-Cygni profile, which complicates a reliable profile fitting of the wings of Ly$\beta$ (Lebouteiller et al.\ 2006). 

Older starbursts ($\gtrsim10$\,Myr) are dominated by B stars. Such stars have particularly strong He\1\ lines in the optical, and overall a more complex FUV spectrum because the ionization degree is lower than in O stars. For this reason, photospheric H\1\ lines becomes prominent (Valls-Gabaud 1993). 
As an illustration, the H\1\ lines from the Lyman series in the BCD Mark\,59 show "V-shaped" wings typical from B stars photospheres (Thuan et al.\ 2002). Pox\,36 shows a similar pattern, but while Thuan et al.\ (2002) got around the problem by considering an artificial continuum to fit the profile of the H\1\ lines in Mark\,59, we decided to model the stellar absorption in Pox\,36 (Sect.\,\ref{sec:stellar}). 

\clearpage

\begin{figure*}
\centering
\includegraphics[angle=0,width=18cm, height=20cm,clip=true]{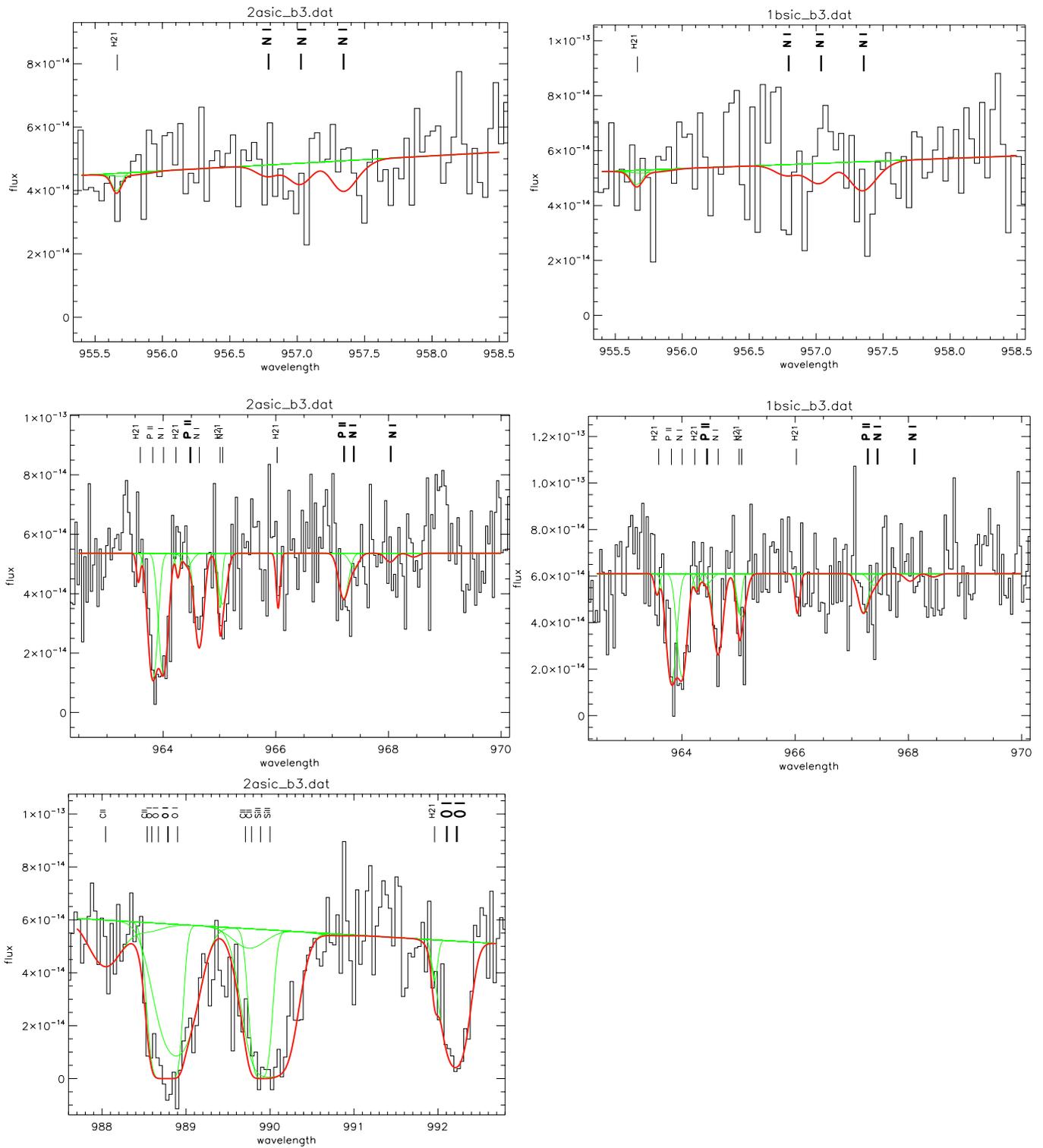}
\caption{Results of the profile fitting. The histogram shows the data. The thin profiles indicate individual absorption-lines while the thick profile indicates the total absorption. Lines labelled in bold arise in Pox\,36 while the others arise in the Milky Way. Wavelength is expressed in \AA, while flux is expressed in ergs\,s$^{-1}$\,cm$^{-2}$\,\AA$^{-1}$.
\label{fig:fitsa}}
\end{figure*}

\clearpage

\begin{figure*}
\centering
\includegraphics[angle=0,width=18cm, height=20cm,clip=true]{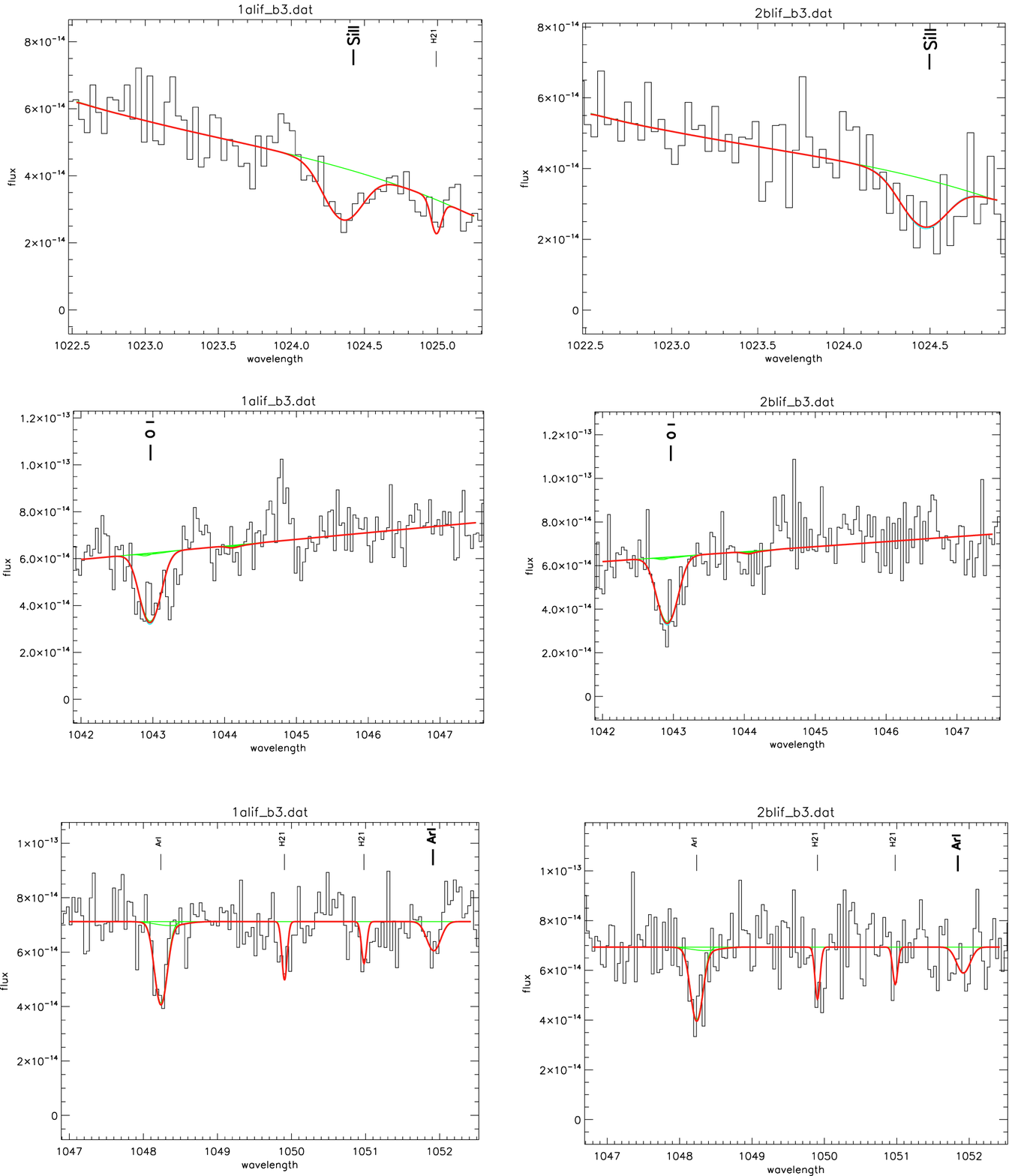}
\caption{See Fig.\,\ref{fig:fitsa} for the plot description.
\label{fig:fitsb}}
\end{figure*}

\clearpage

\begin{figure*}
\centering
\includegraphics[angle=0,width=18cm, height=20cm,clip=true]{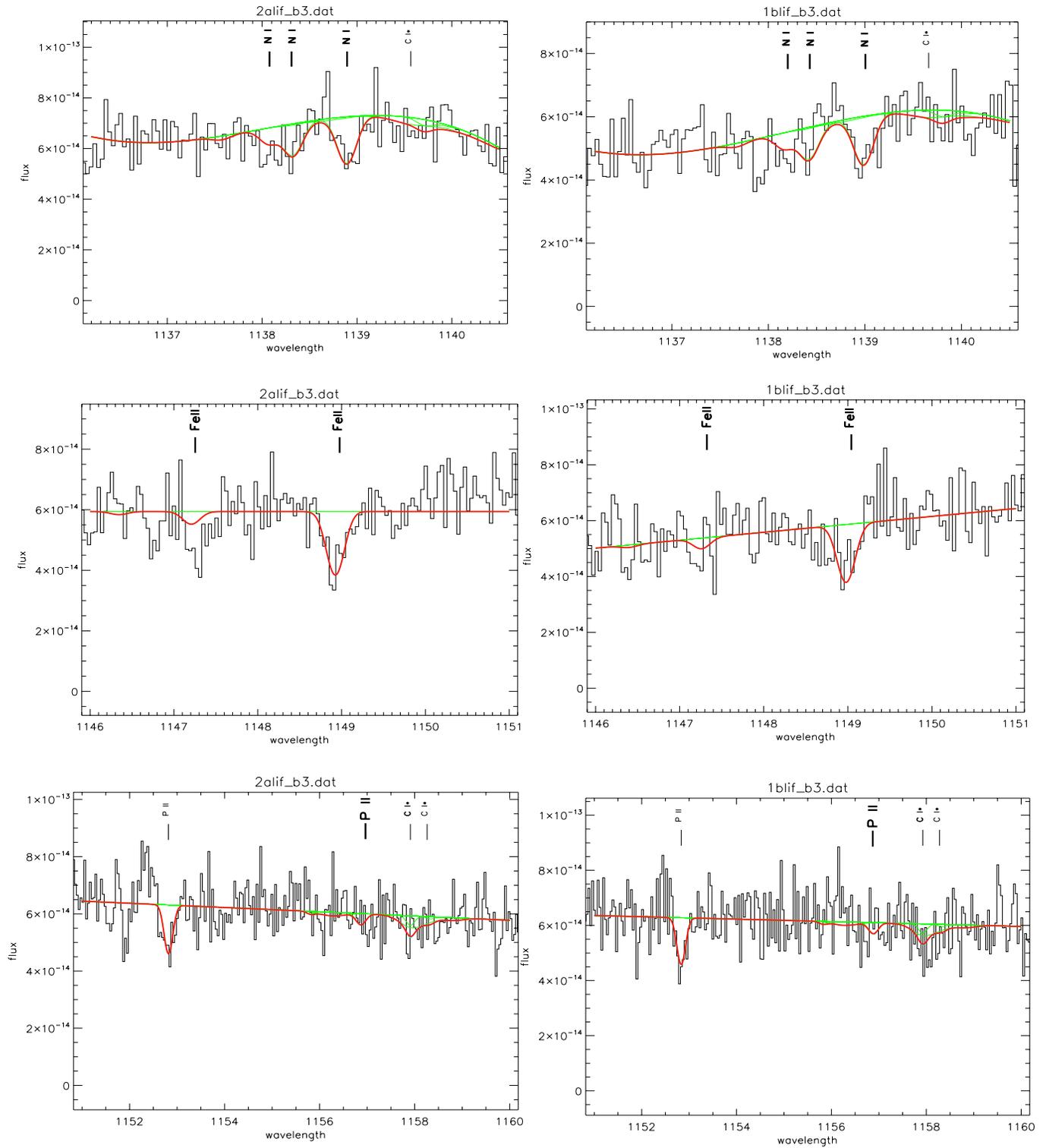}
\caption{See Fig.\,\ref{fig:fitsa} for the plot description.
\label{fig:fitsc}}
\end{figure*}

\clearpage

\section{Column densities}\label{sec:fitting}

In order to infer chemical abundances, we derived column densities based on the line profiles. We adjusted the profiles of the metallic lines using the observed spectrum (Sect.\,\ref{sec:metals}). The H\1\ column density was calculated from the Ly$\beta$ line in the spectrum corrected for stellar absorption (Sect.\,\ref{sec:stellar}).

\subsection{Metals}\label{sec:metals}

\subsubsection{Single component analysis}\label{sec:single}

For the profile fitting, we first made the usual assumption of a single line of sight intersecting a medium with uniform properties (namely with constant radial and turbulent velocities, and column densities). The actual situation can be more complex, with multiple lines of sight toward the many massive stars and H\2\ regions within the galaxy (Sect.\,\ref{sec:multi}). 

The method to derive column densities from \textit{FUSE} line profiles is discussed in H{\'e}brard et al.\ (2002). Profiles are adjusted using the \textit{Owens} routine (Lemoine et al.\ 2002) through a minimization of the $\chi^2$ between the model and the observations. 
All the lines could be fitted with Voigt profiles.
We considered that all the neutral species within Pox\,36 share the same turbulent and radial velocities. 
Those parameters  as well as the column density of each 
species are free to vary. The local continuum around the lines of interest was approximated by a polynomial of degree 0 to 3. The continuum level and shape are also free parameters. The LSF width (Sect.\,\ref{sec:observations}) has little influence on the column density determination. This is principally due to the fact that the lines providing constraints on the column density are weak. The column density determination of all the species considered in this study varies by less than $0.1$\,dex when forcing the LSF width to vary between 20\kms\ and 45\kms. Other LSF widths result in unsatisfactory fits based on the $\chi^2$ value. Lines were fitted  simultaneously for the different observing channels to benefit from the data redundancy allowed by \textit{FUSE}. 
The fitted lines are those in Table\,\ref{tab:lines}. Final fits are presented in Figs.\,\ref{fig:fitsa}, \ref{fig:fitsb}, and \ref{fig:fitsc} while derived column densities are given in Table\,\ref{tab:cds}.
For consistency, we also derived column densities of metals using the spectrum corrected for stellar absorption (Sect.\,\ref{sec:stellar})

\begin{table}
\centering
\begin{minipage}[t]{\columnwidth}
\caption[]{Column densities. Errors are given at 2\,$\sigma$.}
\label{tab:cds}\renewcommand{\footnoterule}{} 
$$
\begin{array}{p{0.2\linewidth}ll}
\hline
\noalign{\smallskip}
Specie & N_{\rm obs}\footnote{$N_{\rm obs}$ is the column density calculated using local spectral continuum from the observation.}   &   N_{\rm norm}\footnote{$N_{\rm norm}$ is the column density calculated using the spectrum corrected from stellar absorption (Sect.\,\ref{sec:stellar}).}  \\
               & ({\rm cm}^{-2})  & ({\rm cm}^{-2}) \\
               \noalign{\smallskip}
\hline  
\noalign{\smallskip}
N\1\ & 14.20^{+0.12}_{-0.14} &           13.96^{+0.22}_{-0.33}      \\
O\1\ & 15.58^{+0.12}_{-0.09} &    15.52^{+0.10}_{-0.09}              \\ 
Si\2\ & 15.06^{+0.19}_{-0.28} &         15.00^{+0.13}_{-0.13}            \\
P\2\ & 12.78^{+0.22}_{-0.32} &            12.94^{+0.16}_{-0.21}              \\
Ar\1\ & 13.18^{+0.19}_{-0.29} &    13.10^{+0.31}_{-0.79}             \\
Fe\2\ & 13.98^{+0.11}_{-0.13} &         13.97^{+0.13}_{-0.18}  \\
\noalign{\smallskip}
\hline
\end{array}
$$
\end{minipage}
\end{table}

\subsubsection{Presence of saturated unresolved components}\label{sec:multi}

In addition to the simple case of a single line of sight intersecting a neutral cloud with uniform properties (Sect.\,\ref{sec:single}), we also consider 
the case of a collection of diffuse clouds producing multiple unresolved absorption components. Such clouds can have different physical conditions and chemical properties. It is commonly accepted that a complex absorption structure along a single line of sight usually does not introduce systematic errors on the column density determination (Jenkins 1996). However, the presence of multiple lines of sight inevitably complicates the analysis. The most obvious effect is the wavelength smearing due to the distribution of the lines of sight across the aperture (Sect.\,\ref{sec:observations}).

The unusually weak interstellar lines in Pox\,36 may indicate that unresolved absorption components $-$ if present $-$ are not saturated. In such a case, the equivalent width of the integrated absorption line is simply the sum of the equivalent widths of the individual components, although the contribution from lines of sight toward stars with different UV brightnesses might not scale linearly (Labasque \& Boiss{\'e} in preparation). We check for the presence of possible saturated components by following the method described in Lebouteiller et al.\ (2006). We derive the maximum column density the narrow component ($b=2$\kms) in the center of a line can have before becoming saturated. The resulting values are given in Table\,\ref{tab:lines} and should be compared to the column densities inferred from the global profiles (Table\,\ref{tab:cds}). It shows that O\1\ and Si\2\ lines could be strongly saturated if narrow components are present, that the Fe\2\ line could be saturated, but that N\1, P\2, and Ar\1\ lines are only close to saturation. Note that this represents only the worst-case scenario since intervening clouds might have in fact larger velocity dispersion. Nevertheless, these results emphasize the reliability of the N\1, P\2, and Ar\1\ column density determinations.

\subsection{Hydrogen}\label{sec:stellar}

The interstellar H\1\ lines are strongly contaminated by absorption from stellar atmospheres (Sect.\,\ref{sec:stellar_cont}). This can be best seen for the high-order lines of the Lyman series in Fig.\,\ref{fig:spectrum} which display typical "V-wings". The derivation of the interstellar H\1\ column densities is done in three steps:
\begin{itemize}
\item estimate the stellar H\1\ absorption using the high-order Lyman lines (e.g., Ly$\epsilon$) whose wings are purely stellar in origin,
\item correct Ly$\beta$ (whose wings are partly stellar and interstellar) by subtracting the stellar H\1\ absorption,
\item estimate the interstellar H\1\ profile using the corrected Ly$\beta$ line.
\end{itemize}

We used the TLUSTY FUV absorption model spectra for O and B stars of Lanz \& Hubeny (2003, 2007) to reproduce the spectral continuum of Pox\,36. We chose to model the stellar spectrum of Pox\,36 with a single stellar population, which we will show in Sect.\,\ref{sec:startemp}, is a safe assumption.
TLUSTY models are available for several temperatures and metallicities. We considered only main-sequence stars ($\log g=4.0$) and have included the effects of stellar rotation. The average rotational velocities have been taken from Kippenhahn \& Weigert (1990) and Lang (1992).
Finally, it must be noted that the velocity dispersion of stars within Pox\,36, or between individual H\2\ regions, is negligible before the stellar rotation velocity. To compare with the observed spectra, we normalized the latter using a spline fit for the continuum, and the synthetic spectra using the featureless continuum provided by TLUSTY.

\subsubsection{Stellar temperature}\label{sec:startemp}

The wings of the stellar H\1\ lines are strongly dependent on temperature. The best constraint is given by the high-order lines of the Lyman series since the interstellar contribution in these lines is saturated and they have no damping wings. Moreover, H\1\ lines are not affected by metallicity so that they are not prone to the degeneracy between temperature and metallicity.

\begin{figure}
\centering
\includegraphics[angle=0,scale=0.38,clip=true]{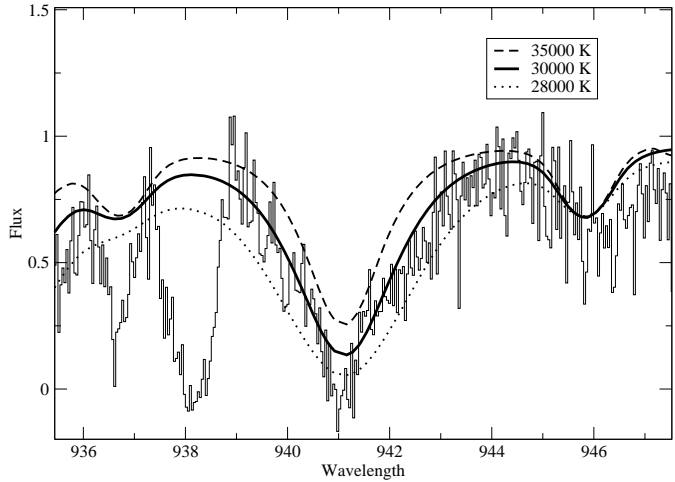}
\caption{The Ly$\epsilon$ line profile gives a strong constraint on the photosphere temperature of the dominant stellar population. Models are drawn for a metallicity of 1/1000\,Z$_\odot$, but the profile of stellar H\1\ lines is unchanged for varying metallicities.
\label{fig:epsilon}}
\end{figure}

Given the redshift of Pox\,36 and the resulting contamination by metallic lines from the Milky Way (see full spectrum in Fig.\,\ref{fig:spectrum}), Ly$\epsilon$ is the cleanest H\1\ line for use to constrain the stellar temperature. Its line profile gives a temperature of 30\,000\,K (Fig.\,\ref{fig:epsilon}), corresponding to a B0 stellar population. The uncertainty on this temperature determination is smaller than the steps allowed by the models ($<2\,500$\,K). Are other stellar types present in large quantities in Pox\,36 ? Hotter stars would contribute narrow H\1\ absorption lines, not modifying appreciably the "V-shaped" wings of the high-order Lyman lines. As for the wings of the low-order Ly$\beta$ line, they are dominated by interstellar absorption. Cooler stars would produce more prominent "V-wings" which are not seen in the profiles of the high-order Lyman lines. We conclude that the stellar populations responsible for the stellar H\1\ profile are mainly composed of B0 stars, with a possible small amount of hotter stars.

Based on the surface flux given by the models and assuming a distance of 20\,Mpc (Table\,\ref{tab:prop}), 
the observed FUV flux of Pox\,36 implies that the equivalent number of B0 stars is $\sim300$. 
This is the number of unextincted foreground stars.

\subsubsection{Stellar metallicity}\label{sec:starmet}

Lines from ionized metals (C\3-\4, N\3-\5, Si\4-\5, ...) arise in the stellar atmospheres of O and B stars. These lines can also contaminate the interstellar spectrum although their effect is much weaker as compared to the stellar H\1\ lines. One expects that the metallicity of young stars to be similar to that of the H\2\ gas (see e.g., Venn et al.\ 2001; Lee, Skillman \& Venn 2006), being equal to $\sim1/5$\,Z$_\odot$ (Table\,\ref{tab:prop}) in the case of Pox\,36.

\begin{figure}
\centering
\includegraphics[angle=0,width=9cm, height=7cm,clip=true]{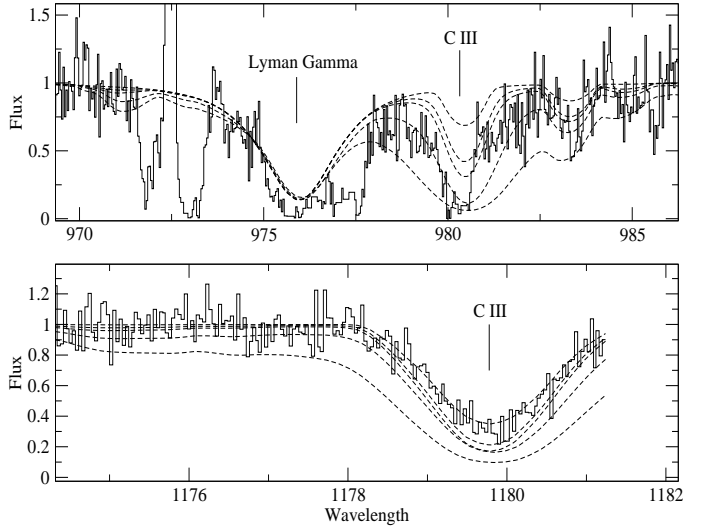}
\caption{The profile of the C\3\ lines at 977.02\,\AA\ (top) and at 1175.6\,\AA\ (bottom) is strongly affected by the metallicity. The models are from bottom to top for $Z=1/2$, $1/10$, $1/50$, $1/100$, and $1/1000$\,Z$_\odot$.
\label{fig:meta}}
\end{figure}

\begin{figure*}
\centering
\includegraphics[angle=0,width=17cm,height=6cm,clip=true]{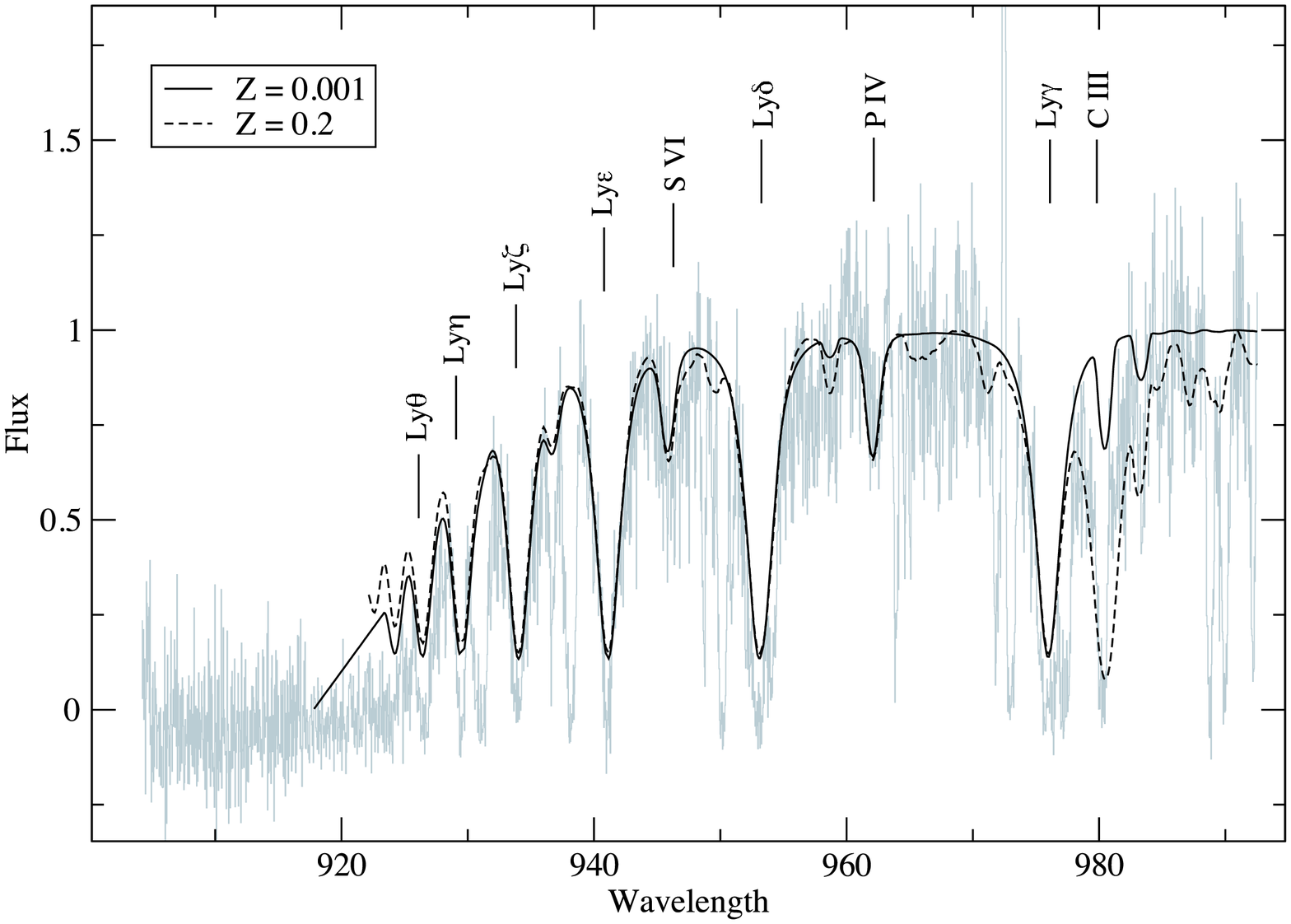}\\
\includegraphics[angle=0,width=17cm,height=6cm,clip=true]{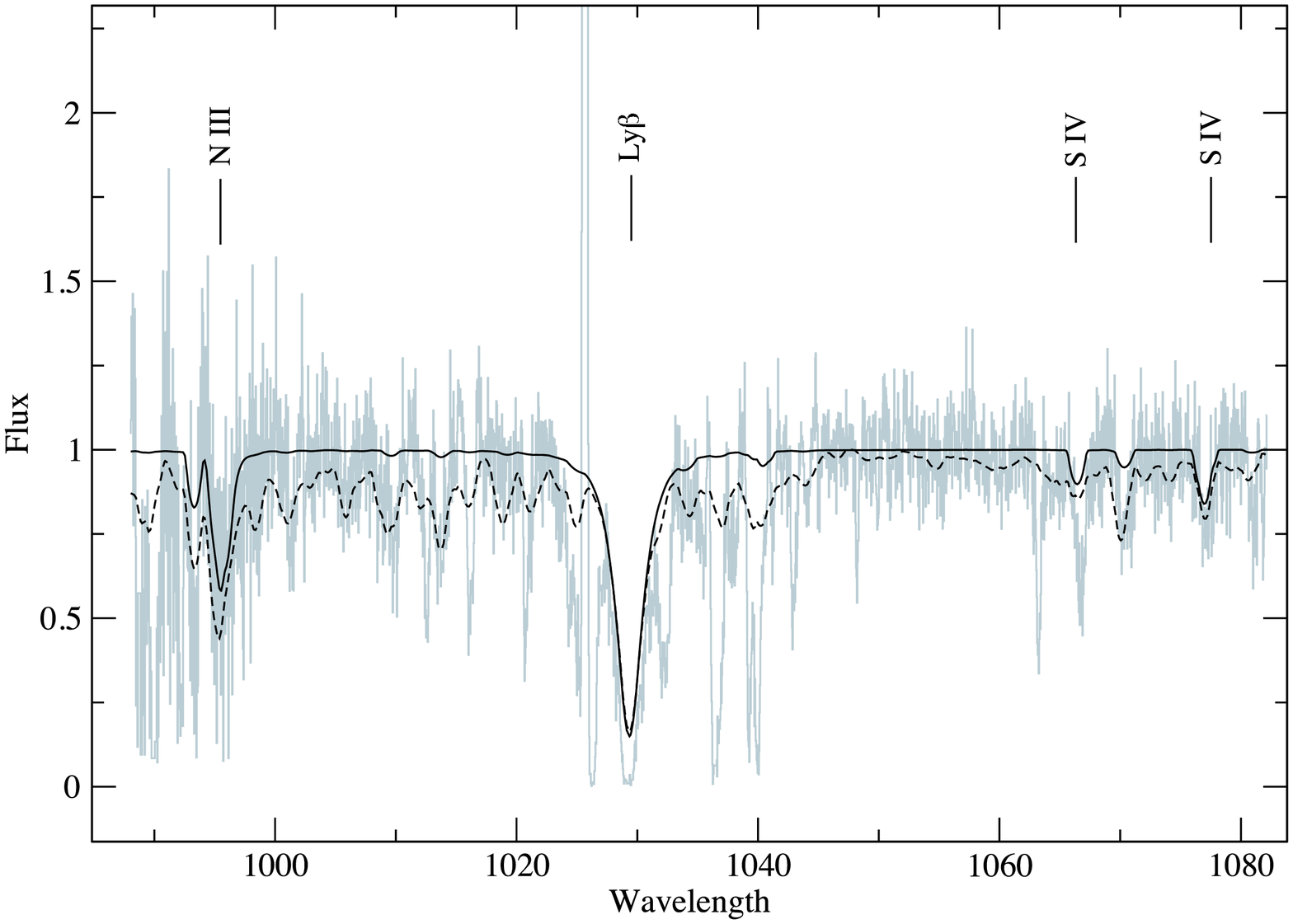}\\\includegraphics[angle=0,width=17cm,height=6cm,clip=true]{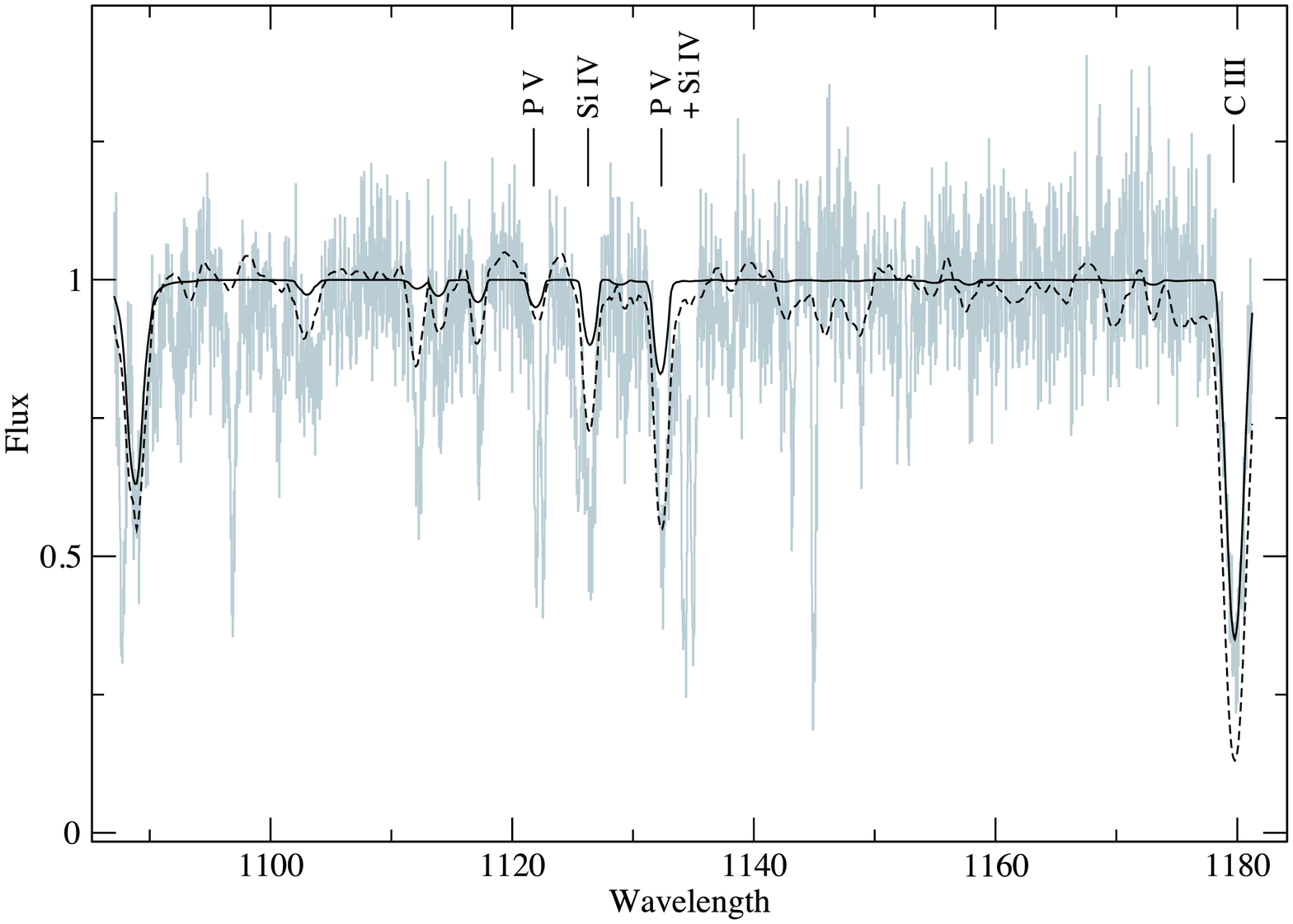}
\caption{The stellar models for $Z=1/1000$\,Z$_\odot$ (solid line) and $1/5$\,Z$_\odot$ (dashed line) are superimposed on the normalized observed spectrum (gray histogram). The most prominent stellar lines are labelled.
\label{fig:stellarobs}}
\end{figure*}

We computed TLUSTY models for a range of metallicities so they can be used to compare with observations. We use the prominent C\3\ lines at 977.02\,\AA\ and at 1175.6\,\AA\ to derive an estimate of the stellar metallicity. The 977.02\,\AA\ line is a resonance line which can be either interstellar or stellar. It gives an upper limit to the stellar absorption. On the other hand, the 1175.6\,\AA\ multiplet is not resonant and is purely stellar. Based on the two C\3\ lines, we find that the observations agree best with a metallicity of 1/1000\,Z$_\odot$ (Fig.\,\ref{fig:meta}). This model fails however to reproduce the stellar absorption lines between 1110\,\AA\ and 1120\,\AA\ (Fig.\,\ref{fig:stellarobs}). Inversely, the 1/5\,Z$_\odot$ model reproduces some stellar lines while overestimating other lines (notably around $\sim1145$\,\AA).

Thus it seems difficult to estimate the stellar metallicity by comparing the observations with the stellar lines predicted by models.
This is of little importance for the interstellar H\1\ column density determination, but it may 
prevent a reliable estimate of the continuum for measuring metal column densities. 
In practice, the stellar lines are broad enough that they don't affect significantly the profiles of the narrow interstellar lines. We decided to use the 1/5\,Z$_\odot$ TLUSTY model and to rederive column densities of the metals using the normalized stellar absorption-corrected spectrum. The results are given in the last column of Table\,\ref{tab:cds}. They are consistent within the errors with those derived from spectra without correction for stellar metallic absorption, with perhaps the exception of the N\1\ column density. Part of the small difference is probably due to systematic errors introduced in the normalization of the observations. Because the two determinations agree relatively well, we decide to use from now on the metal column densities derived from the observations without correction for stellar metallic absorption (Sect.\,\ref{sec:abundances}).

\subsubsection{Interstellar H\1\ column density}\label{sec:ism_hi}

The interstellar H\1\ content is mostly constrained by the damping wings of  the Ly$\beta$ line, since higher-order lines are saturated.
In order to correct for the stellar absorption, we use the TLUSTY stellar model. We chose a metallicity of $1/1000$\,Z$_\odot$ instead of $1/5$\,Z$_\odot$ to avoid uncertainties on the metallic stellar lines which we already saw cannot reproduce well the observations (Sect.\,\ref{sec:starmet}). This is a safe choice for deriving interstellar H\1\ column density since the H\1\ line profiles are independent of metallicity. In practice, our results on the H\1\ column density do not depend on the metallicity chosen for the stellar models. 

\begin{figure}
\centering
\includegraphics[angle=0,scale=0.52,clip=true]{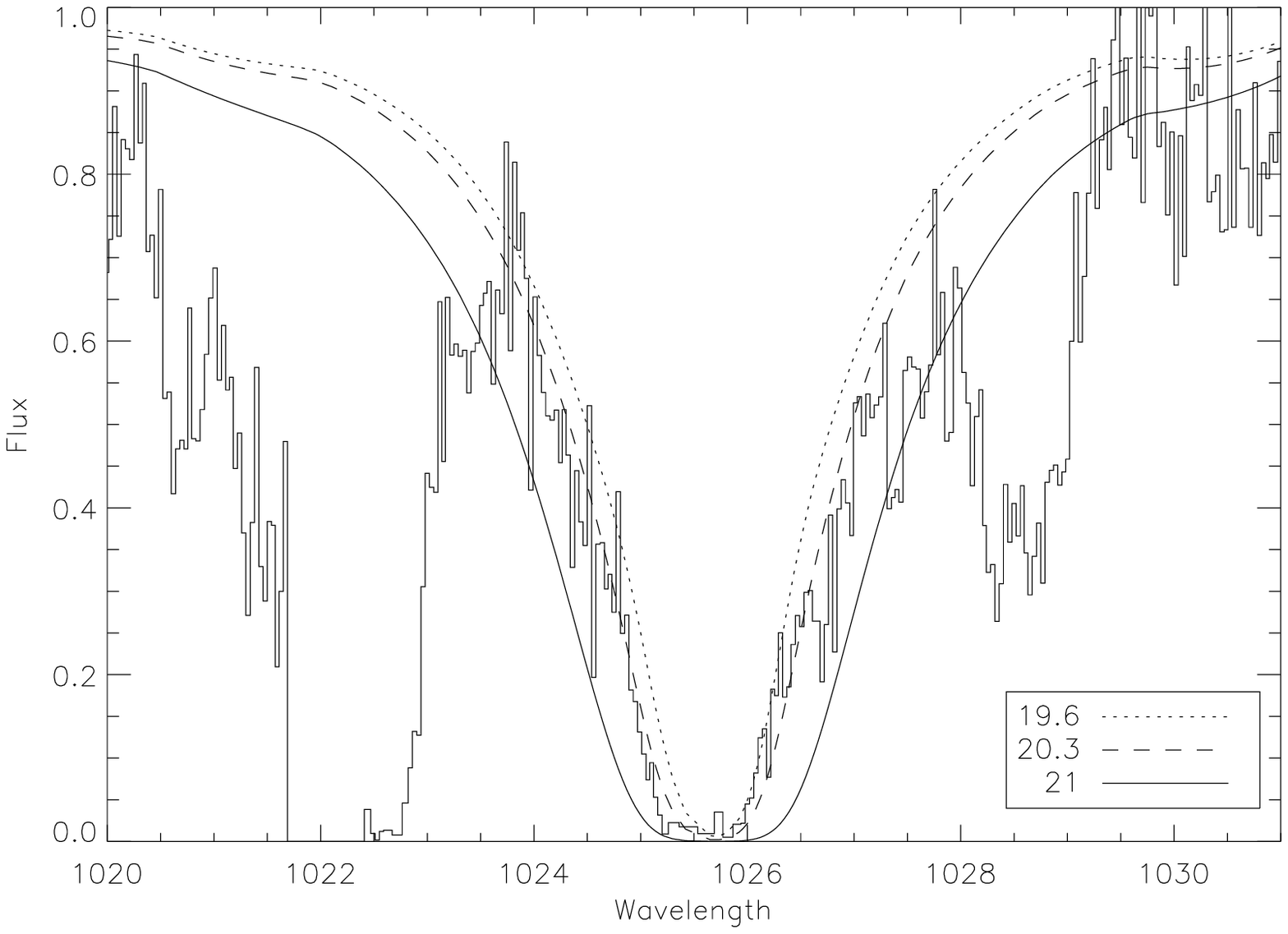}
\caption{The observed Ly$\beta$ profile is compared to synthetic profiles combining the stellar model ($30\,000$\,K) and interstellar H\1\ profiles with various column densities. 
\label{fig:lybeta}}
\end{figure}

\begin{figure}
\centering
\includegraphics[angle=90,scale=0.39,clip=true]{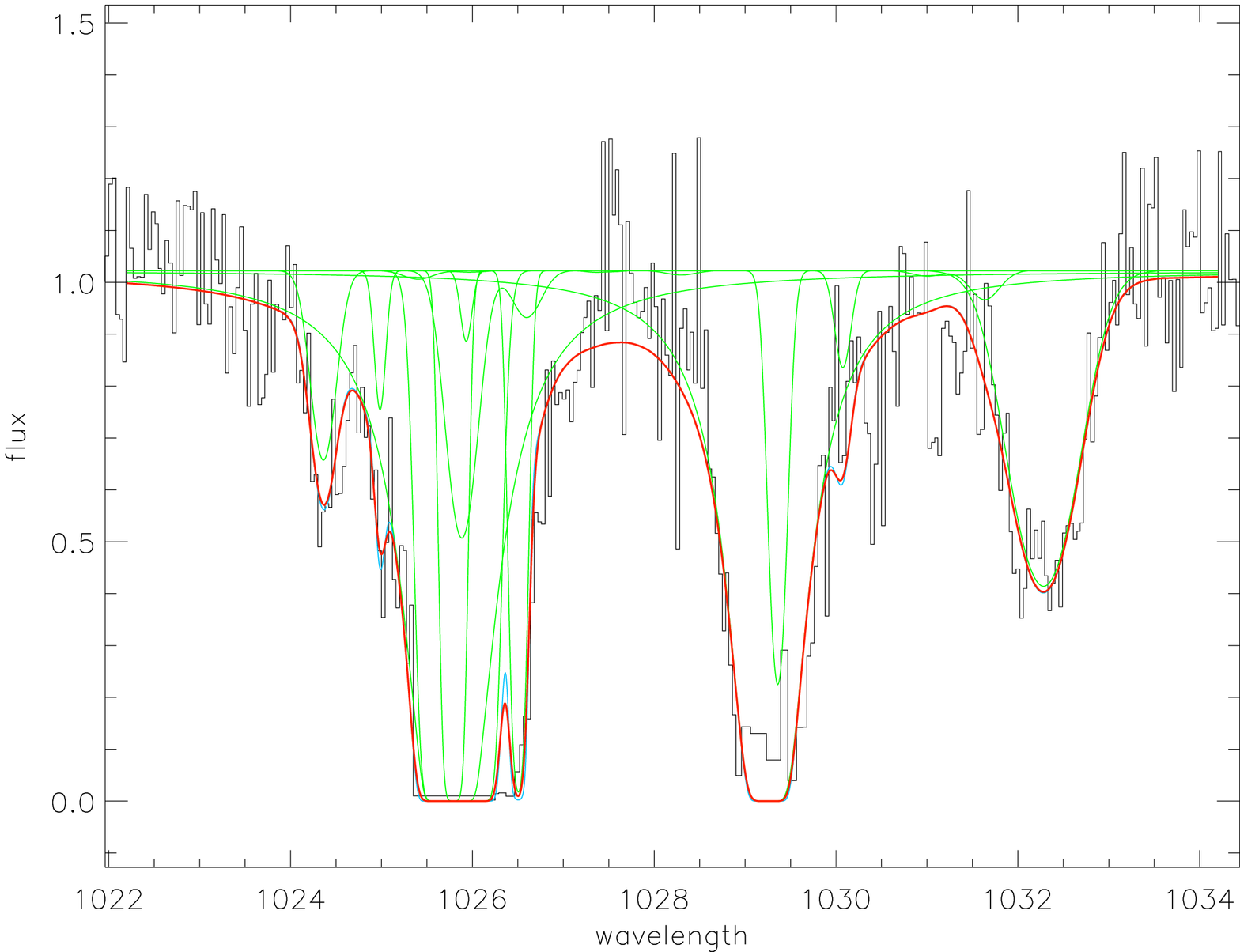}
\includegraphics[angle=90,scale=0.39,clip=true]{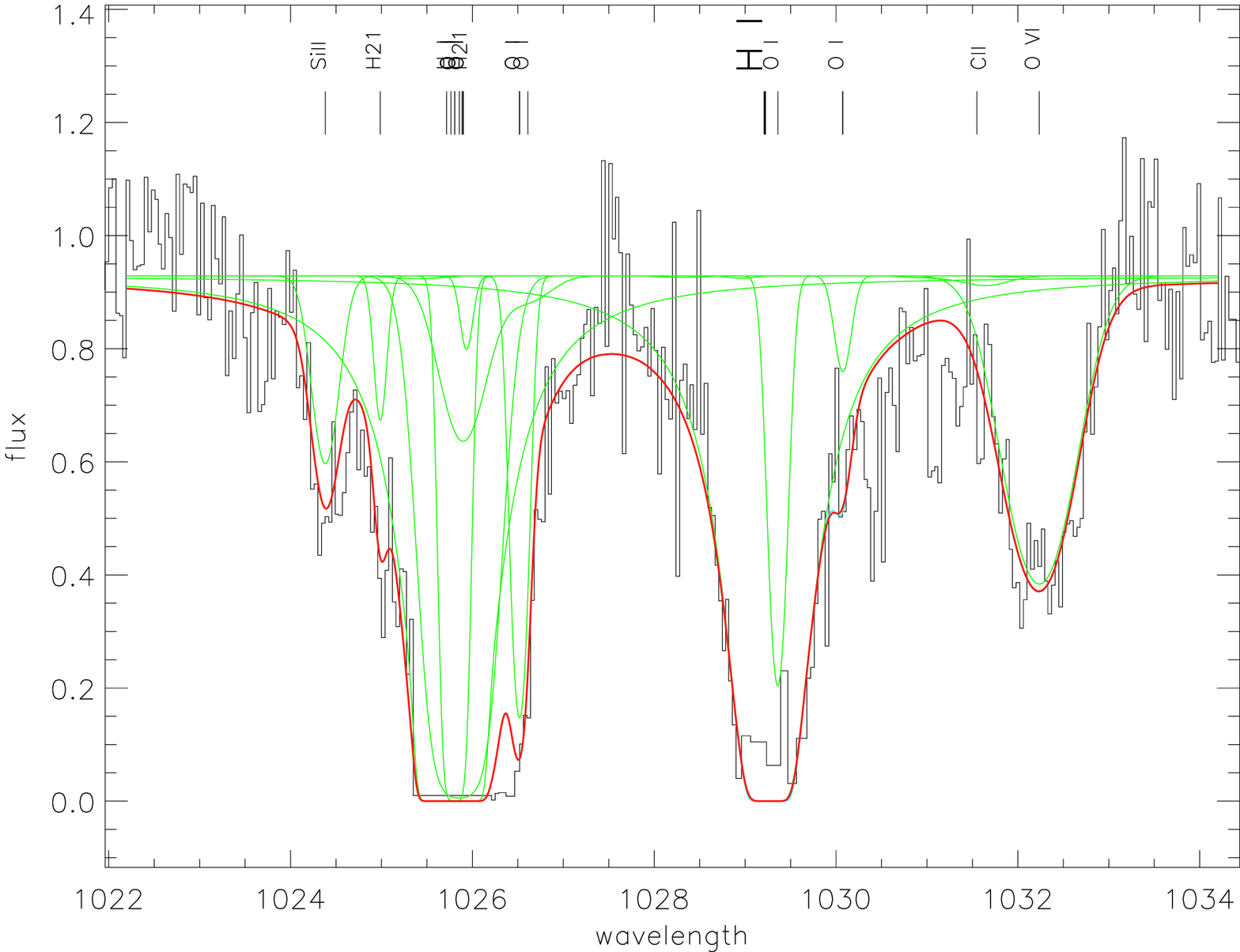}
\includegraphics[angle=90,scale=0.39,clip=true]{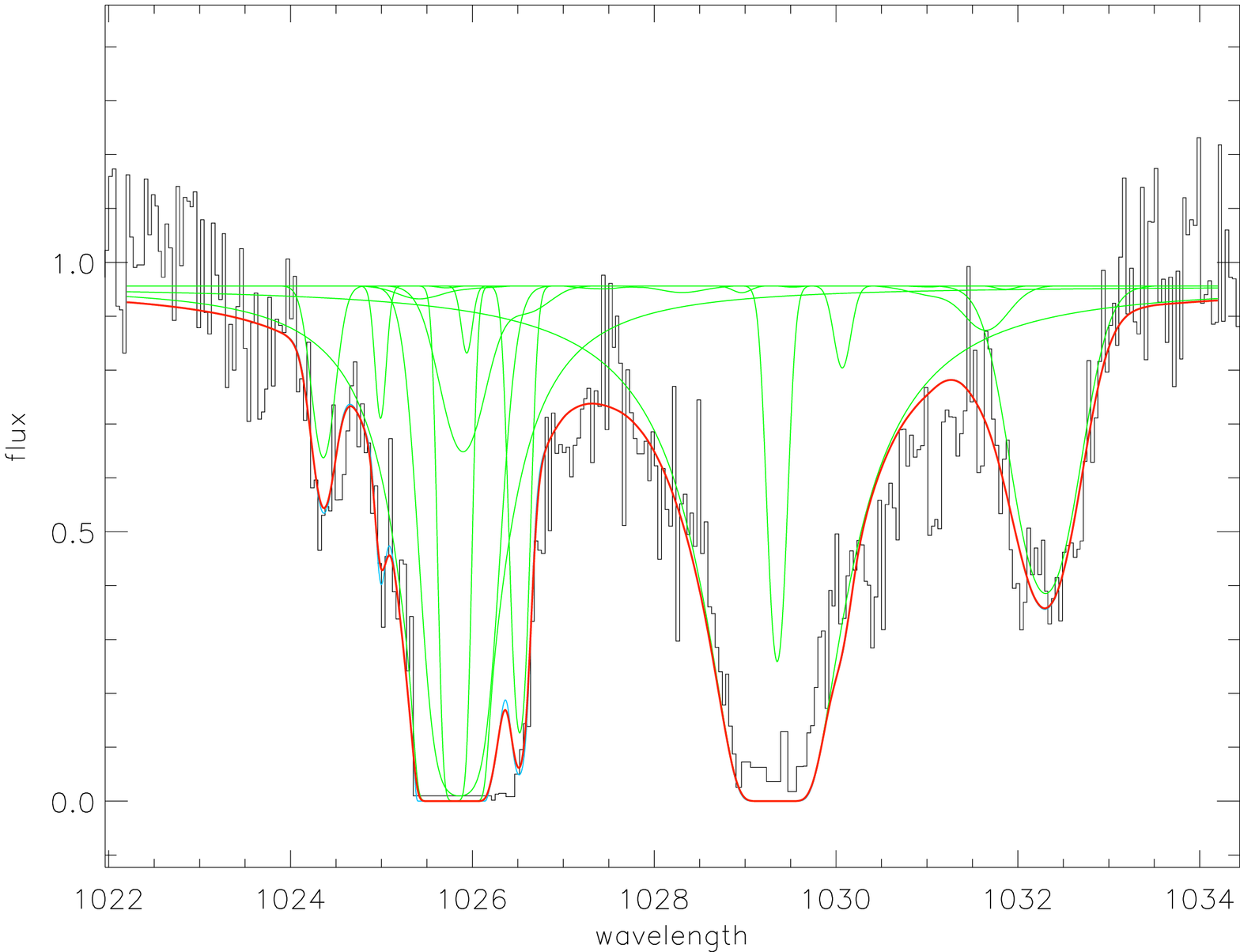}
\caption{Profile fitting of Ly$\beta$ in Pox\,36 at the observed wavelength of $\approx1029.4$\,\AA. The spectra are corrected for stellar absorption using models at $29\,000$\,K (top), $30\,000$\,K (middle), and $32\,500$\,K (bottom). The corresponding line in the Milky Way can be seen at the nominal wavelength of $\approx1025.72$\,\AA.
\label{fig:fitshi}}
\end{figure}

We thus built the interstellar spectrum by removing the stellar contribution. In Fig.\,\ref{fig:lybeta}, the observations are compared to synthetic profiles taking into account stellar and interstellar absorptions. More general cases with various stellar temperatures and lines other than Ly$\beta$ can be found in the appendix \ref{ap:profiles}. The best model agrees with $\log N$ (H\1) $\approx 20.3$. The final fit, after removal of the stellar absorption is shown in Fig.\,\ref{fig:fitshi}. The most likely column density is $\log N$ (H\1) $= 20.28\pm0.06$. The uncertainty is only statistical and does not include systematic errors on the stellar model. We studied the influence of the stellar model on the H\1\ column density by considering models at $29\,000$\,K and $32\,500$\,K. The fits can be found in Fig.\,\ref{fig:fitshi}. The H\1\ column density for the $29\,000$\,K models is $20.18\pm0.08$. For such a temperature, even Ly$\beta$'s profile is almost independent of the interstellar contribution. In fact, the interstellar H\1\ profile for $28\,000$\,K is already saturated without any damping wings, and the H\1\ column density is not constrained ($\log N$ (H\1) $<20.1$). For $32\,500$\,K, the H\1\ column density is $20.66\pm0.06$. In that case, the higher temperature results in narrow stellar lines, and the interstellar contribution, and damping wings are clearly seen. Since a temperature of $32\,500$\,K is clearly too high to reproduce correctly high-order Lyman lines (Sect.\,\ref{sec:startemp}), we consider that the corresponding H\1\ column density should be regarded as a conservative upper limit. It must be noted that the fit quality (given by the $\chi^2$) is almost identical for the 3 temperatures we tested. The reason is that Ly$\beta$ is not a strong constraint on the stellar temperature contrary to Ly$\epsilon$ (Sect.\,\ref{sec:startemp}). We use hereafter the interstellar H\1\ column density $\log N$ (H\1) $= 20.3\pm0.4$ with conservative error bars which include uncertainties on the best stellar model, on the stellar model itself, and on the interstellar profile fitting.

Is this determination in agreement with radio observations? Assuming the neutral gas to be optically thin and a distance of 20\,Mpc, the integrated H\1\ radio flux (Table\,\ref{tab:prop}) gives a mass of $M({\rm HI})=8.1\times10^8$\,M$_\odot$. If the H\1\ distribution is uniform, the column density we derived implies that the H\1\ diameter is 13\,kpc. Based on the image of Fig.\,\ref{fig:pox36}, the angular diameter of the galaxy in the optical is close to $30''$, which at the distance of Pox\,36 corresponds to $\approx3$\,kpc, i.e., a factor $\approx4$ smaller than the H\1\ extent. This is consistent with H\1\ observations of BCDs (e.g., Thuan \& Martin 1981). We conclude that the H\1\ column density measured in the FUV is in agreement with radio observations.

\section{Chemical abundances}\label{sec:abundances}

In this section we derive the chemical abundances from the column densities. Abundance determinations are mostly affected by ionization corrections and by element depletion on dust grains.

\subsection{Ionization structure}\label{sec:ionization}

\begin{table}

\centering
\begin{minipage}[t]{\columnwidth}
\caption{Chemical abundances. [X/H] is defined as $\log$ (X/H) - $\log$~(X/H)$_\odot$, where $\log$~(X/H)$_\odot$ is the solar abundance. Solar abundances are from Asplund et al.\ (2005).}
\label{tab:abs}
\renewcommand{\footnoterule}{} 
    $$ 
         \begin{array}{p{0.3\linewidth}lll}
\hline
            \noalign{\smallskip}
Element (tracer) & \log (X/H) & [X/H]_n\footnote{Abundance in the neutral gas (this study).} & [X/H]_i\footnote{Abundance in the ionized gas (Izotov \& Thuan 2004).}    \\
            \noalign{\smallskip}
\hline  
            \noalign{\smallskip}
N (N\1) &  -6.10\pm0.42 & -1.88\pm0.42 & -1.31\pm0.06 \\
O  (O\1) &  -4.72\pm0.41 & -1.38\pm0.41 & -0.61\pm0.05 \\
Si (Si\2) & -5.24\pm0.47  & -0.75\pm0.47 &  \\
P (P\2) & -7.52\pm0.47 &  -0.88\pm0.47 &   \\
Ar (Ar\1) & -7.12\pm0.47 & -1.30\pm0.47 & -0.42\pm0.06 \\
Fe (Fe\2) & -6.32\pm0.42 & -1.77\pm0.42 & -1.62\pm0.11 \\
            \noalign{\smallskip}
\hline
\end{array}
$$
\end{minipage}
\end{table}

The abundance of an element is usually estimated from its primary ionization state and by applying ionization correction factors. We expect that all elements with ionization potentials larger than that of hydrogen (13.6\,eV) are in a neutral state in the H\1\ gas. This is the case for N, O, and Ar. On the other hand, Si, P, and Fe are mostly found as single-charged ions with negligible fractions of neutral atoms.  

Some nitrogen and argon could be ionized in places where hydrogen is neutral. This is a consequence of the large photo-ionization cross-section of N\1\ and Ar\1\ together with the penetration of UV photons in low-density partly-ionized regions of the ISM (Sofia \& Jenkins 1998). We expect the ionization correction factor for N\1\ to be much smaller than for Ar\1. Our results suggest in fact that they are both negligible (Sect.\,\ref{sec:results}).

Inversely, ionization corrections for Si, P, and Fe may be needed if the singly-ionized ions exist in the ionized gas of H\2\ regions and contribute to the observed FUV absorption lines. For the giant H\2\ region NGC\,604, Lebouteiller et al.\ (2006) found that only Si\2\ and P\2\ require significant ionization corrections (0.34\,dex and 0.15\,dex resp.; see also Aloisi et al.\ 2003 in I\,Zw\,18). Thuan et al.\ (2002) estimated the amount of Fe\2\ in the ionized gas of Mark\,59 to be 20\%\ of the total iron abundance derived from the absorption-lines. Therefore, we could expect ionization corrections on the same order to exist in Pox\,36. We neglect them in the following discussion because it has only little influence on our results.

We finally use N\1, O\1, Si\2, P\2, Ar\1, and Fe\2\ to derive chemical abundances (Table\,\ref{tab:abs}). It must be reminded that the H\1\ column density is larger than the metal column densities by several orders of magnitude, so that weak absorption components could contribute to the damped H\1\ line Ly$\beta$ while the corresponding component in metal lines are below the detection limit (see H{\'e}brard et al.\ 2003 for the effect in D/H determinations). This would lower the interstellar abundance determinations as compared to their true values (Sect.\,\ref{sec:results}).

\subsection{Depletion on dust grains}\label{sec:depletion}

The refractory elements Si and Fe are likely to be locked on dust grains and thus depleted from the diffuse phase. To compare the abundances of the neutral and ionized gas, we need to take into account potential depletion effects.

It is difficult to assess the depletion of elements in the ionized gas of Pox\,36 since only one refractory element, Fe, has been observed in this phase. The use of iron as a probe of depletion is complicated by the fact that it might be released in the ISM on larger timescales. In the Milky Way, it is thought that about 2/3 of the iron in the ISM was produced by SNe type Ia involving low-mass stars. Rodriguez \& Esteban (2004) and Izotov et al.\ (2006) observed that the [O/Fe] ratio increases with metallicity in emission-line galaxies, being consistent with depletion of Fe on dust grains, and with iron production in SN type II in those galaxies.
The [O/Fe] ratio in the ionized gas of Pox\,36 is $\textrm{[O/Fe]}_i=1.12\pm0.12$, i.e., much larger than the mean value of $0.40\pm0.14$ in the ionized gas of BCDs (Izotov \& Thuan 1999). Accordingly, this should imply a significant iron depletion on dust grains in this phase.

The dust content in the neutral gas of BCDs is unfortunately largely unknown. 
FUV observations only select lines of sight with little dust content. Vidal-Madjar et al.\ (2000) have explained the absence of diffuse H$_2$ in the BCD IZw18 by  by the low abundance of dust grains (together with the low H\1\ volume density and the hard radiation field). To date, no diffuse H$_2$ has been detected in other BCDs, including Pox\,36 (Sect.\,\ref{sec:ismcomp}). Therefore, we expect Fe to be less depleted in the neutral gas of Pox\,36 than in the ionized gas. This is the case: the [O/Fe]$_n$ in the neutral gas of Pox\,36 is $0.39\pm0.16$, much lower than the value in the ionized gas ($1.12\pm0.12$), but consistent with the mean value in the ionized gas of BCDs (Izotov \& Thuan 1999). We conclude that iron is nearly not depleted on dust grains in the neutral gas of Pox\,36. Furthermore, since silicon is usually less depleted than iron (e.g., Sofia, Cardelli \& Savage 1994; Savage \& Sembach 1996), our silicon abundance determination should also be little affected by dust depletion.

\section{Results}\label{sec:results}

Based on the metal abundances of Table\,\ref{tab:abs}, it is clear that the neutral gas in Pox\,36 is not pristine: it has already been enriched with heavy elements. This is of course expected since Pox\,36 is more metal-rich than I\,Zw\,18 and SBS\,0335--052 where no pristine gas has been found either (Aloisi et al.\ 2003; Lecavelier et al.\ 2004; Thuan et al.\ 2005). The origin of metals in the neutral phase is however unclear. The presence of old stellar population in the great majority of BCDs (e.g., Aloisi et al. 1999; {\"O}stlin 2000) leads to the conclusion that previous enrichment could have occured in previous starburst episodes or through low-level star-formation over the galaxy's history. Comparison with the chemical composition of the ionized gas in the H\2\ regions provides an interesting diagnostic to understand the enrichment process of the neutral gas.

\subsection{Comparison with the ionized gas of the H\2\ regions}\label{sec:compa}

Table\,\ref{tab:abs} shows that N is underabundant by a factor $\approx4$ in the neutral gas as compared to the ionized gas, while both O and Ar are underabundant by a factor $\approx6$ and $\approx8$ respectively. The relatively good agreement between the deficiency factors of N, O, and Ar in the neutral gas indicates that the abundances of N and Ar are probably well determined (Sect.\,\ref{sec:ionization}).
The iron abundance appears to behave differently, being the same within the errors in the two gaseous phases. However, if we correct [O/Fe] in the ionized gas of Pox\,36 for dust depletion (Sect.\,\ref{sec:depletion}), the iron abundance becomes:
\begin{equation}
\textrm{[Fe/H]}^\textrm{*}_i \sim \textrm{[Fe/H]}_i + \textrm{[O/Fe]}_i - 0.40,
\end{equation}
giving [Fe/H]$^\textrm{*}_i$ $\sim-0.9$, about 7 times higher than the value in the neutral gas. Thus Fe is underabundant in the neutral gas by the same factor as N, O, and Ar.

Can this metal deficiency in the neutral gas be attributed to an overestimate of the H\1\ column density? Considering the metal column densities of Table\,\ref{tab:cds}, the neutral gas would have the same metallicity as the ionized gas if the H\1\ column density is on the order of $\log N$ (H\1) $\approx 19.6$. This is inconsistent with the observations since such a low column density does not reproduce the wings of Ly$\beta$ (Fig.\,\ref{fig:lybeta}). We conclude that the  discontinuity between the metallicity of the neutral and ionized gas is not driven by uncertainties on the H\1\ column density determination.

Oxygen and argon are the best metallicity tracers available in the ionized gas, and agree with a metallicity of $\approx1/5$\,Z$_\odot$ in this phase. Considering these 2 elements, we find that the metallicity in the neutral gas is $\approx1/35$\,Z$_\odot$, i.e., a factor $\sim7$ below the value in the ionized gas. Our results for Pox\,36 are similar to those obtained for other BCDs studied by \textit{FUSE} (Sect.\,\ref{sec:global}): although the neutral gas of Pox\,36 is not pristine, it is less chemically evolved than the ionized gas in the H\2\ regions.

\subsection{Pristine gas pockets}\label{sec:pristine}

It is possible that pockets of pure H\1\ lie in the halo, while only a fraction of the H\1\ has been enriched close to the starburst region. This would dilute the chemical abundances inferred from integrated lines of sight by adding H\1\ without any metallic counterpart. 
In Pox\,36, the H\1\ has a radial velocity slightly lower than that of the metals (Sect.\,\ref{sec:ismcomp}). 
The same result was observed in IZw18 by Aloisi et al.\ (2003) where the H\1\ has a velocity of $753\pm6$\kms\ as compared to $765\pm10$\kms\ for the metals. If the offset is real (i.e., not driven by systematic uncertainties on the wavelength calibration), this suggests that part of the H\1\ could lie indeed in regions without any metals. This would imply that the ISM of the galaxy is in fact not well mixed, with much lower abundances toward the external regions. The offset is however too small to assert unambiguously the presence of metal-free gas.

\subsection{Enrichment process}\label{sec:calculs}

We now calculate the amount of metals produced and released by the present starburst episode in order to understand the metal enrichment of the neutral gas. The stellar mass formed is derived from the UV luminosity, following Lequeux et al.\ (1981). We assume an initial mass function slope of $-1.5$ and a burst age of 10\,Myr (see Sect.\,\ref{sec:stellar_cont}). We find that the stellar mass is $2.5\times10^6$\,M$_\odot$.
Yields were taken from Meynet \& Maeder (2002), assuming a metallicity of 0.2\,Z$_\odot$ and a mass interval of $[2-120]$\,M$_\odot$. We find that an oxygen mass of $M(O)=1.25\times10^{4}$\,M$_\odot$ is eventually released by the starburst. Several scenarios are then possible, depending on the volume extent over which the dispersal and the mixing occurs.

We first assume that all the newly produced oxygen mixes with the ionized gas. Following Kunth \& Sargent (1986), the mass of ionized gas is:
\begin{equation}
M(HII) = \frac{N m_p}{n_i \beta},
\end{equation}
where $N$ is the number of ionizing photons, $m_p$ is the proton mass, $n_i$ is the electronic and ionic density, and $\beta$ is the recombination coefficient. From the H$\beta$ luminosity (Table\,\ref{tab:prop}), we find $N=4.4\times10^{50}$\,s$^{-1}$. 
We calculate $\beta=2.482\times10^{-14}$\,cm$^3$\,s$^{-1}$ from the Table of Hummer \& Storey (1987) using the electron density ($60$\,cm$^{-3}$) and temperature ($12\,560$\,K) found by Izotov et al.\ (2004). The mass of ionized gas is then  $M(HII)=2.5\times10^{5}$\,M$_\odot$. If the SNe products mix only with the ionized gas, the metallicity of this phase would reach an additional oxygen enrichment of $12+\log({\rm O/H})=9.64$. This unrealistic value suggests that heavy elements will mix with a larger mass of gas, at larger spatial scales.

Next, we assume that the newly produced oxygen will mix uniformly with the H\1\ gas. Using the H\1\ mass from Table\,\ref{tab:prop}, we find that the neutral gas would be enriched by an additional 5\%\ only ($12+\log({\rm O/H})=5.98$). Therefore, it can be seen that many starburst episodes ($\sim20$) are required for the neutral gas to reach the metallicity level we derived with \textit{FUSE}. 

A possible alternative is that the mixing of new metals might not be uniform as the H\1\ lying close to the stellar clusters could be more enriched than the outskirts which would remain metal-free. In order for a nearby H\1\ gas shell to have the same metallicity as that of the star-forming region (ionized gas), we find that the newly produced oxygen should mix with $7.0\times10^{7}$\,M$_\odot$ of H\1, i.e., $\sim10$\%\ of the total H\1\ in the galaxy. 
This can be seen as the upper limit of the H\1\ volume that can be enriched to reasonable levels.
We conclude that SNe products could mix with a small fraction of the H\1\ gas in Pox\,36 while not 
enriching most of it. This hypothesis implies that pockets of pristine gas could well be present in the galaxy. One way to prove this is to study absorption-lines toward several star-forming regions of a given galaxy, with significantly different foreground gas content. Another way is to analyze the scatter of the relation between abundances in the ionized gas and neutral gas for a sample of BCDs (Sect.\,\ref{sec:chemen}). 

Toward a plausible scenario, it could thus be that the vast majority of the metals released during a starburst episode mixes with a large mass of H\1, increasing only slightly its metallicity, while a small fraction mixes with the star-forming region itself (currently ionized), increasing greatly its metallicity. These metals most likely come from previous episodes of star-formation, with the consequence that   star-formation episodes have to be co-spatial. We consider the simple case in which the metallicity of the neutral phase increases by $f_n m / M(HI)$, where $m$ is the mass of metals released by a starburst and $f_n$ is the fraction of these metals to be mixed within the H\1\ region. Similarly, the ionized gas increases its metallicity by $f_i m / M(HII)$. Over successive starburst episodes, a metallicity offset between the ionized and neutral phases is achieved and maintained if the following condition is met: 
\begin{equation}
f_n = \frac{1}{1 + \alpha \frac{M(HII)}{M(HI)} },
\end{equation}
where $\alpha$ is a parameter depending on the metallicity offset. An offset of 1\,dex $-$ as it is observed in BCDs $-$ implies $\alpha=10$. If, as in Pox\,36, the H\1\ mass is 1000 times larger than the ionized gas mass, the fraction of metals mixing with the H\1\ phase should be $\approx99$\%, while only 1\%\ mixes with the gas of the star-forming region. In conclusion, under the previous condition, the neutral gas can remain metal-poor, compared to the ionized gas, as long as the star formation keeps occurring at the same location. If another burst occurs at another location, one might expect to measure low abundances, but the offset will reappear when the next starburst episode occurs and after mixing is achieved.

\section{Discussion}\label{sec:global}

\subsection{Chemical enrichment of BCDs}\label{sec:chemen}

\begin{figure}
\centering
\includegraphics[angle=0,scale=0.37,clip=true]{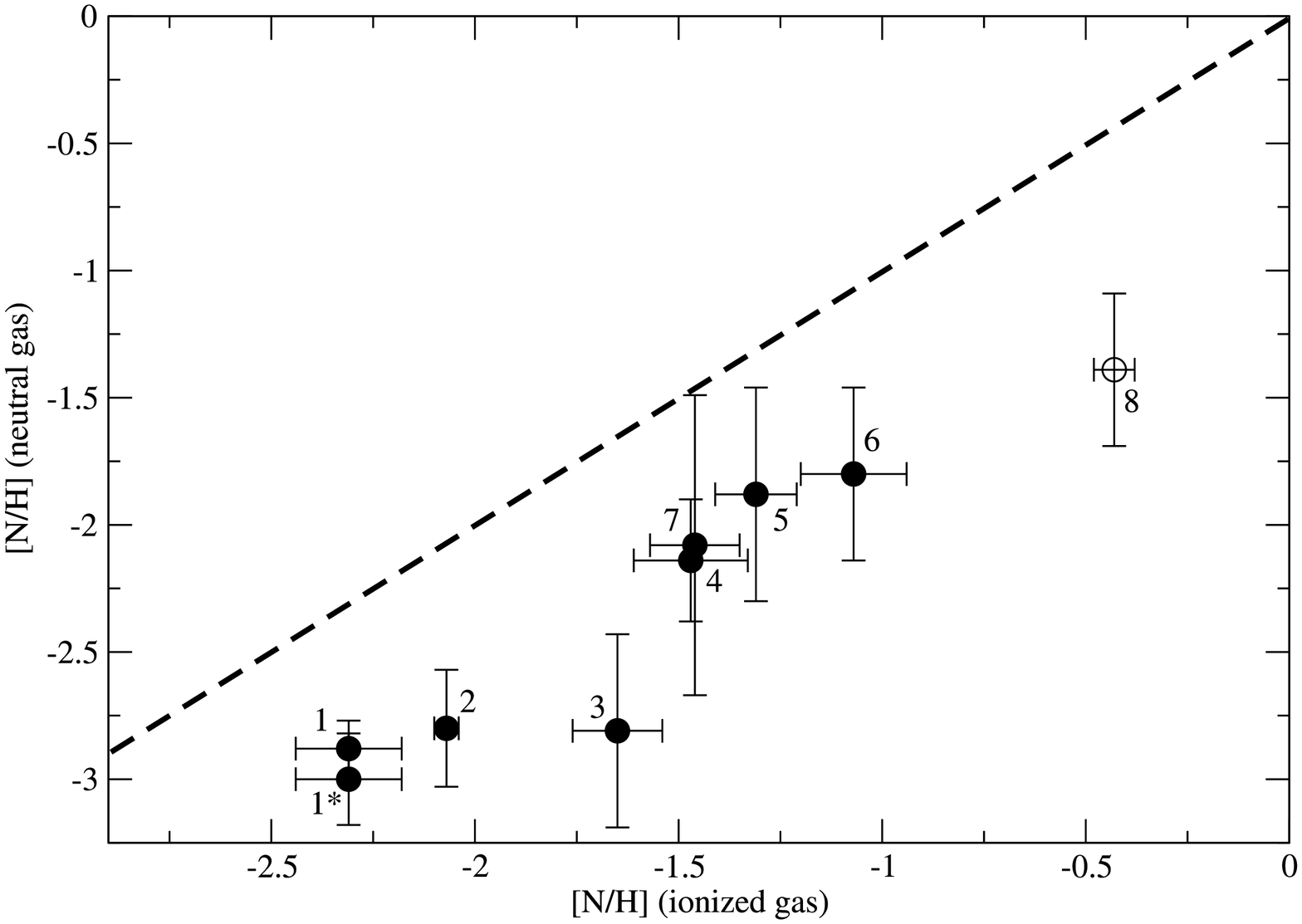}
\includegraphics[angle=0,scale=0.37,clip=true]{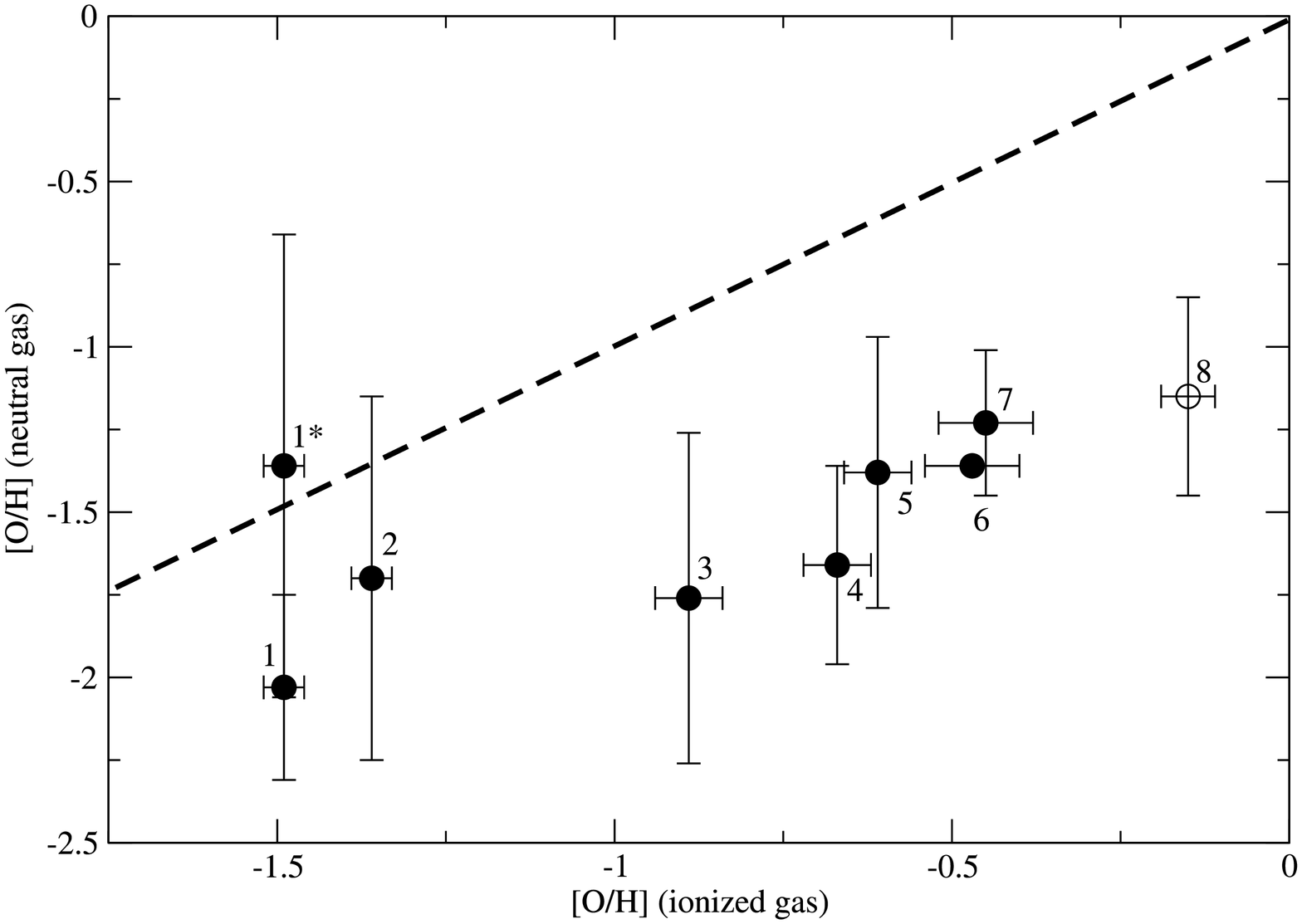}
\includegraphics[angle=0,scale=0.37,clip=true]{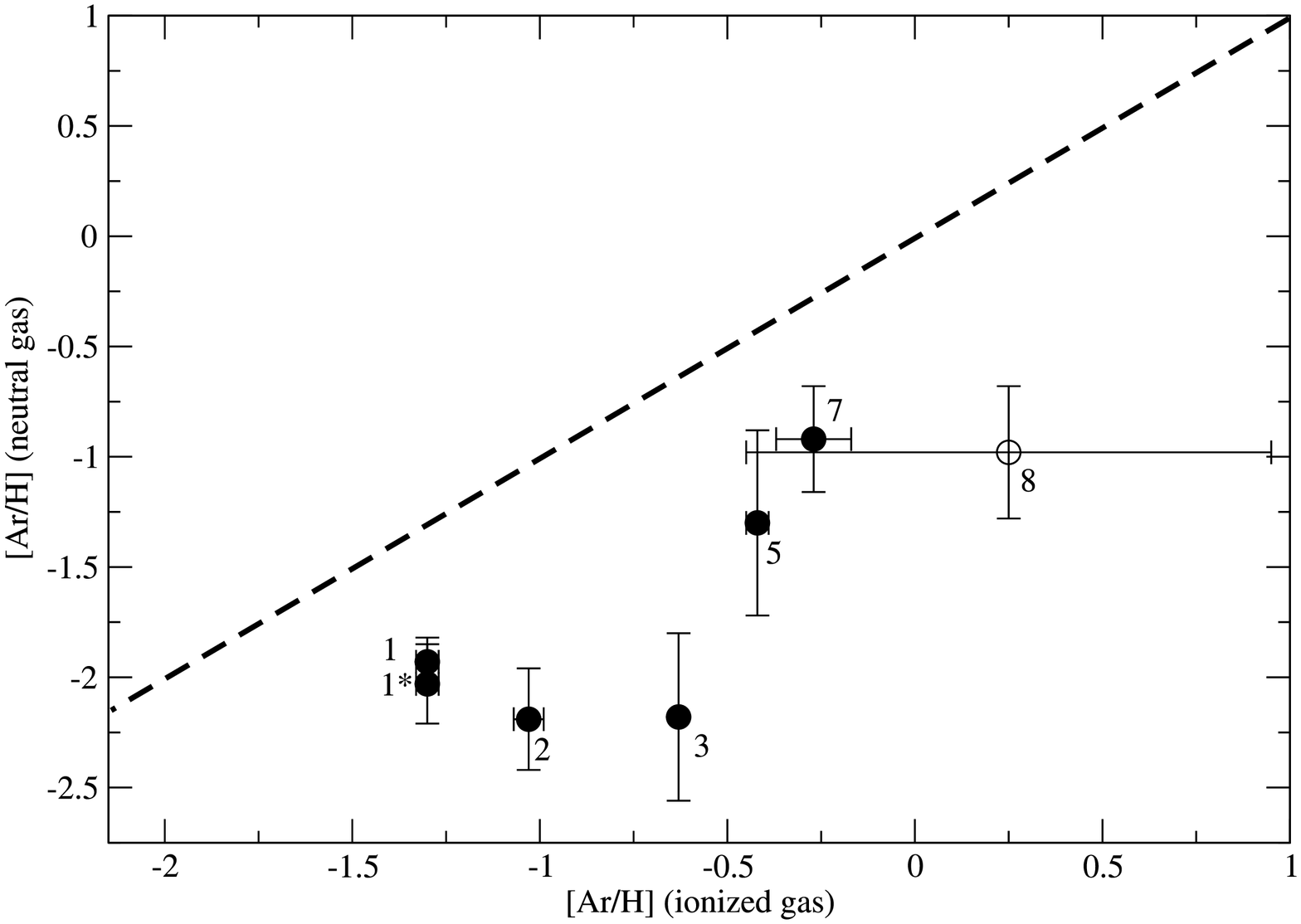}
\caption{The abundances of nitrogen (top), oxygen (middle), and argon (bottom) in the ionized gas and in the neutral gas are compared. 
The dashed line indicates the 1:1 ratio.
Labels: (1*) I\,Zw\,18 (Lecavelier des Etangs et al.\ 2004), (1) I\,Zw\,18 (Aloisi et al.\ 2003), (2) SBS\,0335--052 (Thuan et al.\ 2005), (3) I\,Zw\,36 (Lebouteiller et al.\ 2004), (4) Mark\,59 (Thuan et al.\ 2002), (5) Pox\,36 (this study), (6) NGC\,625 (Cannon et al.\ 2005), (7) NGC\,1705 (Heckman et al.\ 2001), (8) NGC\,604/M\,33 (Lebouteiller et al.\ 2006).
\label{fig:plateau}}
\end{figure}

\begin{figure}
\centering
\includegraphics[angle=0,scale=0.38,clip=true]{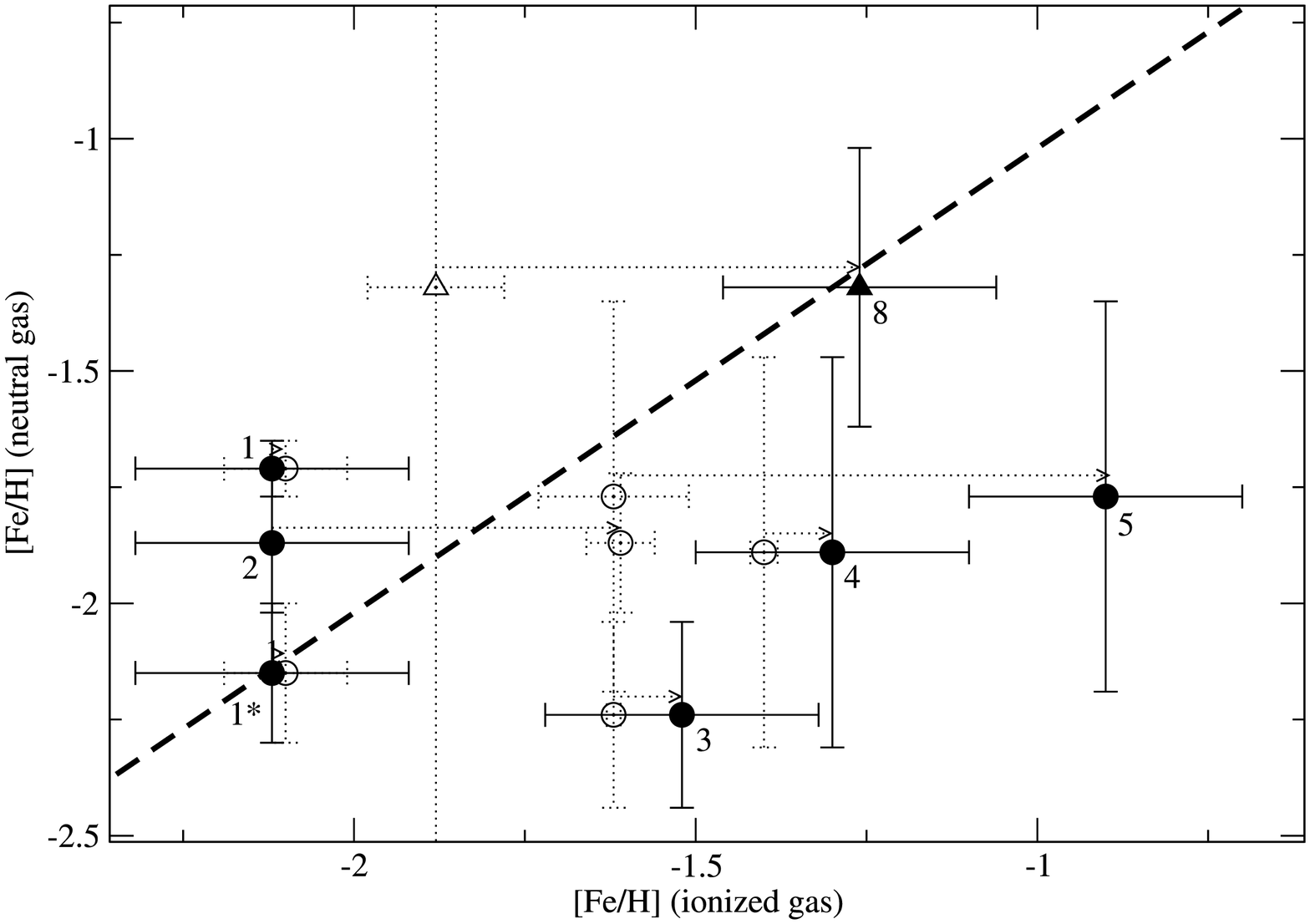}
\caption{Abundances of Fe in the neutral gas and in the ionized gas. Grey points correspond to the [Fe/H] in the ionized gas while black dots include a correction to account for depletion (Sect.\,\ref{sec:results}). See Fig.\,\ref{fig:plateau} for the labels. 
\label{fig:FeH}}
\end{figure}

The \textit{FUSE} BCD sample is now large enough to allow the study of general trends. Are the results concerning Pox\,36 consistent with those of other BCDs? A total of 7 BCDs with various metallicities ($1/50$ to $1/3$\,Z$_\odot$) have now been analyzed, in addition to the high-metallicity giant H\2\ region NGC\,604 in the spiral galaxy M\,33. We plot in Fig.\,\ref{fig:plateau} the elemental abundance [X/H]\footnote{[X/H] is defined as $\log$ (X/H) - $\log$~(X/H)$_\odot$, where $\log$~(X/H)$_\odot$ is the solar abundance from Asplund et al. (2005).} in both the ionized and neutral gas of the BCDs. It can be seen that N, O, and Ar show identical trends, with systematically lower abundances in the neutral ISM than in the ionized gas. Part of the dispersion of [O/H] is probably due to saturation effects which could not be avoided in some objects. Some of the dispersion of [N/H] and [Ar/H] is probably due to ionization corrections. 
Since the trend is similar for N, O, and Ar, the corrections are either of the same order for all objects ($\leq$ 0.5\,dex), or they are negligible, as for Pox\,36 (Sect.\,\ref{sec:results}). The abundance of iron in the neutral and ionized gas of BCDs must be interpreted with more care because of depletion on dust grains (Sect\,\ref{sec:depletion}). 
Figure\,\ref{fig:FeH} shows the relation between the Fe abundances in the two gaseous phases, considering both the observed abundances and the ones corrected for depletion. It can be seen that corrected values agree more with the general trends observed for N, O, and Ar (Fig.\,\ref{fig:plateau}). It is seen in particular that iron depletion affects significantly the ionized gas abundances, especially in SBS\,0335--052, Pox\,36, and NGC\,604.

Examination of [Ar/H] and [N/H] in Fig.\,\ref{fig:plateau} reveals the possible existence of a plateau for the lowest-metallicity BCDs (12+log(O/H)$_i$ $\leq$ 7.8 or $\lesssim1/8$\,Z\,$_\odot$). The data on [O/H]$_n$ are also consistent with the existence of a metallicity plateau at $12+\log {\rm (O/H)}_n \sim 7.0$ ([O/H]$_n = -1.7$), although the error bars are considerably larger.
The presence of a plateau would imply a minimal enrichment of the neutral ISM. 
It could be that the H\1\ reservoir has this minimal metallicity before star-formation ever occurred. Interestingly, 
Telfer et al.\ (2002) found that the intergalactic medium in the redshift range $1.6<z<2.9$ has a metallicity $12+\log ({\rm O/H})$ between $6.7$ and $7.6$, which includes the floor metallicity observed in BCDs. Hence it appears that the neutral and ionized ISM of the lowest metallicity galaxies could well be characterized by a default enrichment rather than by previous star-formation episodes. In opposition, the higher-metallicity objects would be enriched by many starburst episode. Further observations and analysis of BCDs with metallicities lower than $\sim1/8$\,Z\,$_\odot$ will be needed to confirm the existence of this plateau.

At higher metallicities $\gtrsim1/5$\,Z$_\odot$, the abundances of the neutral gas and ionized gas seem to be well correlated. This is the case for Mark\,59, Pox\,36, NGC\,625, and NGC\,1705. The giant H\2\ region NGC\,604 appears to extend the trend to nearly solar metallicities. The neutral gas abundances are systematically lower than those of the ionized gas. Nitrogen, oxygen, and argon in the neutral gas are underabundant by a factor $\sim0.9$\,dex, i.e., nearly 10, as compared to the ionized gas.

If we dismiss the idea of a plateau and consider instead that the points in Fig.\,\ref{fig:plateau} show a positive correlation with significant scatter from the lowest to the highest metallicity objects, then the situation could be consistent with a dilution by metal-free gas along the lines of sight. The scatter would then be due to the dilution factor being different for each object. However one would then expect objects with the largest H\1\ column density to display the largest underabundance in the neutral gas. We studied the dependences of the underabundance factor with the H\1\ column density but found no significant correlation.

\subsection{Interpretation}\label{sec:inter}

The lower metallicity observed in the H\1\ regions of BCDs as compared to the ionized gas of their H\2\ regions remains a puzzle. We have discussed many effects that may affect the determination of the neutral gas abundances such as the source extent, ionization corrections, depletion on dust grains, stellar contamination, and line saturation.
The underabundance of a factor 10 found in the H\1\ region of BCDs is however so dramatic that it seems difficult all these effects can make us miss the detection of 90\%\ of the metals. Therefore, this offset appears to be real and we discuss in the following possible physical explanations. While self-enrichment of the H\2\ regions, as proposed by Kunth \& Sargent (1986), naturally explains why metals are more abundant in the ionized gas, this scenario is difficult to reconcile with the homogeneous abundances observed in disconnected H\2\ regions of most dwarf galaxies (e.g., Dufour 1986; Russell \& Dopita 1990; Skillman \& Kennicut 1993; Kobulnicky \& Skillman 1996; Kobulnicky \& Skillman 1997; Legrand et al.\ 2000; Noeske et al.\ 2000). In fact, the observed dispersion is probably dominated by variations of the electron temperature and not by intrinsic chemical inhomogeneities. Much less frequent is the observation of local discontinuities, with only NGC\,5253 (Welch 1970; Walsh \& Roy 1989; Kobulnicky et al.\ 1997; L{\'opez-S{\'a}nchez et al.\ 2007), II\,Zw\,40 (Walsh \& Roy 1993), IC\,4662 and ESO\,245--G05 (Hidalgo-G{\'a}mez et al.\ 2001). Still, the local pollution could be generally explained by Wolf-Rayet winds (Brinchman et al.\ 2008) rather than by SNe explosions.

Based on the gas kinematics in the dwarf irregular NGC\,1705, Heckman et al.\ (2001) suggested that the absorption of neutral species could arise in small dense clumps evolving in a superbubble interior. The swept-up clumps have likely the same metallicity as the ionized gas of the H\2\ region, but more metal-poor gas could lie further away, diluting the integrated abundances. Alternatively, Cannon et al.\ (2005) proposed that the neutral gas probed by FUV absorption lines comes from a low abundance H\1\ halo, while metal-rich ionized gas lies close to the stellar clusters. The metallicity offset observed between the neutral gas and the ionized gas seems to be independent of the chemical age of the galaxy, so that abundances have to be kept high in the star-forming regions and low in the halo over extended periods. This implies that star-formation has to remain localized over the galaxy's history. 

Our results and calculations (Sect.\,\ref{sec:results}) suggest that most of the metals released by a starburst episode could mix with the H\1\ region, only increasing slightly its metallicity, while a small fraction (on the order of 1\%) mixes locally within the star-forming regions. It must be stressed that these metals were produced by previous bursts. In such a scenario, low-level star-formation is not required to explain the metallicity of the neutral phase (except if galactic winds allow most of the metals to escape in the intergalactic medium). Is this scenario consistent with the observation of homogeneous abundances in H\2\ regions of a given star-forming galaxy? The fact that star-forming episodes occur simultaneously indicates that these episodes were probably triggered together, or that they triggered each other over several 10\,Myr. Hence it is in fact reasonable to expect relatively similar abundances. 

The contrasting result of Bowen et al.\ (2005) in the dwarf spiral galaxy SBS\,1543+593 poses a fundamental question: is the metal deficiency of the H\1\ gas specific to BCDs? Although Lebouteiller et al.\ (2006) found that the neutral phase toward NGC\,604 in the spiral galaxy M\,33  is more metal-poor than the ionized gas of the H\2\ region, i.e., being consistent with results in BCDs, the situation might be different because of the presence of high-velocity clouds along the lines of sight. Rotation in spirals provide an efficient mixing throughout the galaxy. Furthermore, star-formation keeps occurring along the spiral arms. In opposition, BCDs have weak velocity field, and star-formation usually occurs in one or several knots (e.g., Papaderos et al.\ 1996). The different mode of star-formation could be responsible for substantial differences in the dispersal and mixing mechanisms. It seems necessary to study the metal abundance in the neutral gas of dwarf irregular galaxies which are not BCDs. Studies of other star-forming objects thus ought to be performed, which will be possible thanks to the interactive database tool designed by D{\'e}sert et al.\ (2006a; 2006b). Analysis is currently in progress. Besides, one of the most complication in the studies of absorption-line spectra of galaxies is the presence of unseen components, possibly saturated. This caveat is usually solved by the use of the weakest observed lines. In that vein, the upcoming installation of the \textit{Cosmic Origin Spectrograph} onboard the \textit{Hubble Space Telescope} will allow the sulfur abundance determination in the neutral phase through the interestingly weak S\2\ $\lambda1256$ multiplet (Pettini \& Lipman 1995). The oxygen abundance will be measured with a better accuracy than with \textit{FUSE} thanks to the O\1\ $\lambda1356$ line (Meyer et al.\ 1998). Finally, the examination of phosphorus lines will confirm its the possible use of phosphorus abundance as a metallicity tracer as suggested by Lebouteiller et al.\ (2005). This is particularly important for the analysis of the \textit{FUSE} database since O\1\ lines are generally saturated or close to saturation. Finally, it must be stressed that the ideal case study for deriving abundances in the neutral gas of BCDs is to observe the ISM toward a quasar line of sight (Kunth \& Sargent 1986). The method was successfully used in the dwarf spiral galaxy SBS\,1543+593 by Bowen et al.\ (2005) but unfortunately, no alignment BCD-quasar has been observed so far. The study of such an alignment would cancel the selection effect in \textit{FUSE} spectra to probe gas toward star-forming regions within the galaxy.

\section{Summary and conclusions}

We have analyzed the FUV \textit{FUSE} absorption spectrum of the BCD Pox\,36. The following results are found:
\begin{itemize}
\item  We detect the interstellar absorption lines of H\1, N\1, O\1, Si\2, P\2, Ar\1, and Fe\2. Pox\,36 appears to be have a particularly low foreground gas content, which allows us to measure column densities with minimal systematic errors due to hidden saturation effects.
\item Metal column densities have been derived from the line profiles. The interstellar H\1\ lines are strongly contaminated by stellar absorption. We have modeled the spectral continuum using TLUSTY models. The best model is constrained by the high-order Lyman line Ly$\epsilon$ and gives a photospheric temperature of $\approx30\,000$\,K (B0 type). After correcting Ly$\beta$ for stellar absorption, we have used the interstellar damping wings to infer a H\1\ column density equal to $20.3\pm0.4$. 
\item The observed FUV continuum level is consistent with $\sim300$ B0 stars. This represents only the unextincted stars.
\item Abundances have been computed from the column densities using the dominant ionization stage. The large [O/Fe] ratio in the ionized gas of Pox\,36 is interpreted as caused by iron depletion onto dust grains. In contrast, the neutral gas does not show any depletion, which is attributed to the low dust content along the observed lines of sight.
\item We find that the neutral gas is metal-underabundant by a factor 5 to 8 as compared to the ionized gas of the H\2\ regions. This corresponds to a metallicity of $\approx1/35$\,Z$_\odot$ in the neutral gas as compared to $\approx1/5$\,Z$_\odot$ in the ionized gas and imply that the neutral gas is less chemically processed.
\item It is shown that the metal deficiency in the neutral gas can be achieved and maintained if most of the metals released by star-formation episodes mix with the H\1\ gas, increasing only slightly its metallicity, while a small fraction (1\%) could mix locally and enrich greatly the ionized gas. The metals should come from previous bursts which occurred around the same location as the present one. 
\item Based on the sample of the seven BCDs studied by FUSE, we have confirmed that there is a systematic underabundance by a factor $\sim10$ of the neutral gas as compared to the ionized gas in the H\2\ regions. Besides, [Ar/H] and [N/H] seem to reach a minimum value for the lowest-metallicity objects ($\sim1/8$\,Z\,$_\odot$), which could be due to a default enrichment.
\item There is no convincing evidence that the neutral gas in BCDs contains pristine gas pockets, i.e., completely metal-free. However, it could be that the neutral gas of BCDs has not been affected yet by the star-formation episodes, and that its metallicity is that of the IGM. 
\end{itemize}

\begin{acknowledgements}
The authors thank A.\ Lecavelier des Etangs, G.\ H{\'e}brard, R.\ Ferlet, and A.\ Vidal-Madjar for their input as to the handling of the data. T.X.T. acknowledges the support of National Science Foundation grant AST02-05785. He thanks the hospitality of the Institut d'Astrophysique de Paris where part of this work was carried out. 
\end{acknowledgements}

\appendix
\section{Synthetic H\1\ profiles}\label{ap:profiles}

We built H\1\ profiles of Ly$\alpha$, $\beta$, $\gamma$, and $\delta$ by summing the optical depths of the stellar models and of synthetic interstellar lines. Profiles were calculated for various column densities and allow one to know, depending on the stellar temperature, which line can be safely used to constrain the interstellar H\1\ column density. From the results of Figs.\,\ref{fig:interH11} and \ref{fig:interH12}, it is clear that:
\begin{itemize}
\item Ly$\alpha$ can be used for any stellar temperature,
\item Ly$\beta$ can be used for $T\gtrsim30\,000$\,K,
\item Ly$\gamma$ and Ly$\delta$ do not provide any constraint, even for high stellar temperatures. This is because these lines are saturated without any damping wings.
\end{itemize}

\begin{figure*}
\centering
\includegraphics[angle=0,scale=0.82,clip=true]{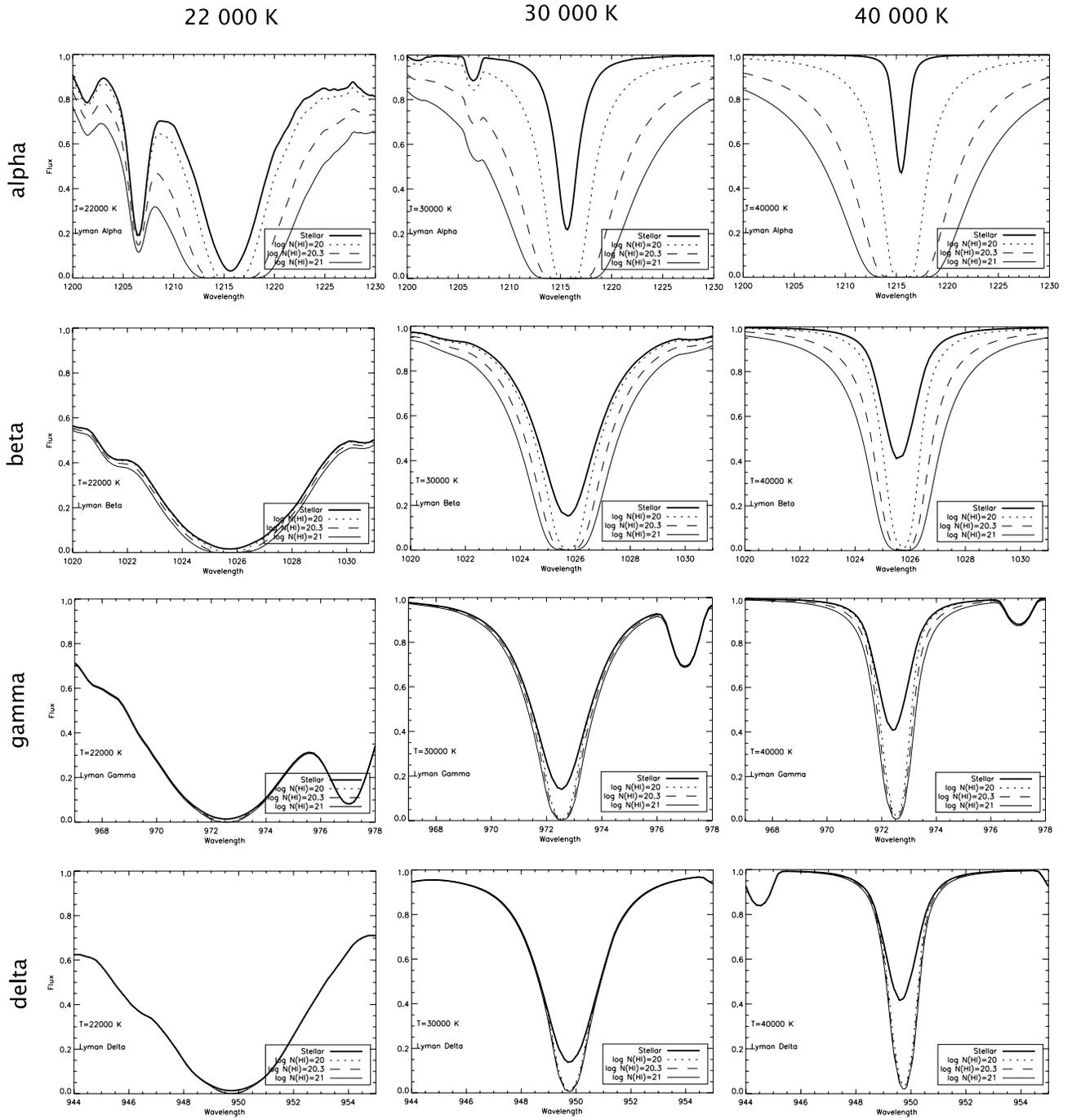}
\caption{Prediction of the total H\1\ profile (stellar and interstellar) of the Lyman lines. All stellar models are for  1/1000\,Z$_\odot$ metallicity except for the model at $22\,000$\,K for which we take 1/5\,Z$_\odot$.
\label{fig:interH11}}
\end{figure*}

\begin{figure*}
\centering
\includegraphics[angle=0,scale=0.82,clip=true]{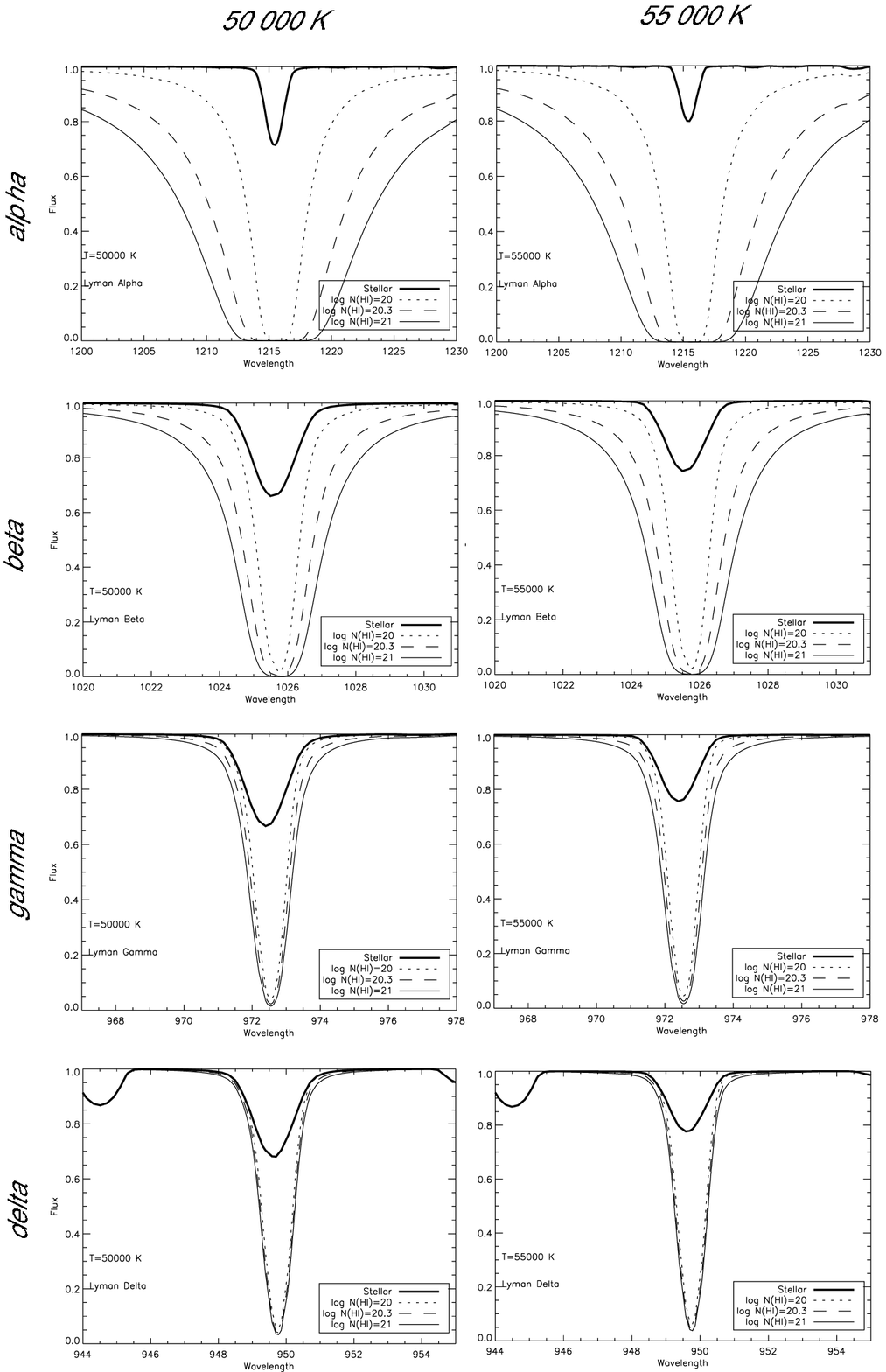}
\caption{See Fig.\,\ref{fig:interH11} for the plot description.
\label{fig:interH12}}
\end{figure*}

\end{document}